# Numerical investigation of erosive wear considering surface degradation using coupled CFD-DEM and a bond model

Vinh D.X. Nguyen[1,2,*], A. Kiet Tieu[1,*], Damien André[3], Hongtao Zhu[1]

[1] School of Mechanical, Materials Mechatronic and Biomedical Engineering, Faculty of Engineering and Information Sciences, University of Wollongong, Northfield Avenue, Wollongong, NSW, 2522, Australia.

[2] Biomolecular Interaction Centre, University of Canterbury, Private Bag 4800, Christchurch 8041, New Zealand.

[3] Institute of Research for Ceramics (IRCER), UMR 7315, F-87000 Limoges, France.

Corresponding Author

[1,2,*] Email: vinh.nguyen@canterbury.ac.nz

[1,*] Email: ktieu@uow.edu.au

**Abstract**

A three-dimensional (3D) simulation model is proposed here to study the erosive wear of structure caused by solid particles, which accounts for the accumulation of surface deformation and degradation during the erosion process. Although there are numerous studies on the erosion, they have been primarily based on semi-empirical equations to predict the material loss and lack of an engineering approach to account for the surface evolution during erosion process, which is proved to be of great important. Our proposed model aims to overcome these challenges. In this approach, a conventional volume of solid object is discretized into a number of discrete spherical elements. A bond network is added to connect neighbouring particles at micro-scale, forming a cohesive structure at macro-scale. It is indeed a combination of particulate flow simulation by coupled Computational Fluid Dynamics (CFD) - Discrete Element Method (DEM) and materials modelling by DEM using a bond model. An extensive study of fluid response in both laminar and turbulent flow is conducted when the traditional continuous rigid object is replaced by a corresponding discrete deformable object composed of bonded spherical particles. The replacement of discrete object allows this model to naturally incorporate the three-way interaction of fluid, structure, and particles. The interaction between fluid and structure at macro-scale is implicitly represented by the interaction between fluid and bonded particles at micro-scale, while material properties and the micro-structure of the solid object are reproduced by properties of bond that account for deformation and fracture. Both resolved and unresolved coupling schemes result in good agreement with the traditional CFD approach. Although the unresolved scheme is less accurate for the near wall streamlines reproduction, it preserves the general flow field compared to the rigid solid wall. It is computationally efficient in the case of large number of particles, hence it is suited to the industrial scale problem. The coupled CFD-DEM using bond model has been demonstrated as an alternative numerical method for investigating the fundamentals of erosion or complex fluid-structure interaction problems where the microstructure of materials is important. However, there exists certain limitations of our model, which will be discussed further in the benchmarks.

**Declarations:**

**Conflicts of interest:** The authors declare that they have no conflict of interest.

**Availability of data and material:** Not applicable

**Code availability:** GranOO, OpenFOAM, LIGGGHTS, CFDEM

## 1. Introduction

Recently, we have seen a substantial amount of research on discrete element method (DEM), particularly in simulating the mechanical behaviour of various types of materials and coupled particle-fluid flow simulation. DEM is an alternative approach to model wear or fracture of materials, while the coupled particle-fluid simulation provides a solution for multiscale and multiphase problem including all types of interaction between fluid-particle, particle-particle and particle-wall. In the continuum mechanics on macroscopic scale, materials such as rock, concrete or steel are usually described as homogeneous continua, and their mechanical properties are often modelled by continuum methods, particularly Finite Element Method (FEM). In various fracture phenomena, the material under consideration often consists of both continuous and discontinuous parts. In order to deal with these problems, Moes, et al. [1] developed the Extended Finite Element Method (XFEM), which added enrichment functions on the nodes where cracks initiate and propagate. However, due to the dependence on mesh quality and sensitivity to the choice of appropriate shape/enrichment function, XFEM faces challenges in handling large deformation, multiple fractures, wear with debris creation and microscopic structure of materials. In certain cases, the microstructure of materials, which consists of a large number of inclusions and mixture of several phases, is too complex to be modelled by a single continuum mechanics law. For such problems, a multi-scale approach plays a crucial role in investigating the practical phenomena across different time and length scales.

Meanwhile, particle-based method is a rapidly developed interdisciplinary research area aimed at gaining insight into the relationship between micro- and macro- properties of materials. Discrete Element Method (DEM) was proposed by Cundall and Strack [2] to study the mechanical behaviour of geomaterials like soil and rock. Unlike FEM and other continuum mechanics-based methods, DEM does not require a continuous mesh to represent the solid object. Instead, the solid is discretized into a collection of discrete particles interacting with each other. In the continuum approach, the material properties are described by stress-strain constitutive laws whereas in DEM, the mechanical behaviour is introduced by force-displacement laws through the contact model. For example, Fig. 1**a, b** and **c** demonstrate the components of a general contact model. DEM has advantages in its ability to naturally account for discontinuities and is therefore a good alternative to the continuum approach for simulating discontinuous phenomena such as fracture. However, DEM has two main drawbacks. Firstly, the classical continuum mechanics laws cannot be directly applied within a DEM formulation. Secondly, due to the spherical shape of particles, the volume between the discrete elements/particles creates an artificial void inside the material. To simulate cemented rocks, Potyondy and Cundall [3] introduced Bonded Particle Model (BPM) by adding a parallel bond (Fig. 1**d**). After that, a wide variety of bond models have been developed and validated to improve the ability of DEM to simulate the continua through practical applications, from homogeneous materials such as ceramics, steel and aluminium to complex multi-physics behaviour like damage of heterogeneous solids like concrete or rock. For instance, André, et al. [4] has used the cohesive beam bond model based on Euler-Bernoulli beam theory to model perfectly elastic materials. Its remarkable applications and validations are the study of brittle fracture and replication of Hertzian cone crack in indentation of silica glass [5]. The fracture phenomena due to thermal effect is also investigated by this method [6]. Furthermore, Le, et al. [7] presented the idea of using DEM with bond model to simulate the composite materials with different materials and structure for matrix and fiber. The early concept of using cohesive plastic beam to model ductile materials was presented by Terreros [8], and subsequently improved by Nguyen, et al. [9] to study typical mechanical phenomena of steel and aluminium. The implementation of cohesive beam bond is introduced in the open source GranOO software [10]. Obermayr, et al. [11] present a new beam bond based on linear finite-element Timoshenko beam element to simulate cemented sand. The applications of DEM using bond models are not only quasi-static phenomena like tension, compression, but also dynamic problems like impact and abrasive wear. In a recent work, André, et al. [12] presents a new approach using Voronoi cells to completely fill the volume space of discrete domains.

In addition to applications in modelling materials, DEM has also been applied to track the particles movement in the particle-fluid flow. Particle laden flow can be observed in a variety of scientific and engineering processes, among others, the study of dispersion of pollutants in environment (air, water, soil), gas fluidization and pneumatic conveying are popular examples. The multiscale and multiphase simulation, combining Computational Fluid Dynamics (CFD) and Discrete Element Method (DEM), is an optimal choice for studying those phenomena, in which the carrier fluid is considered the continuum phase and the particles are the discrete phase. The coupling between CFD and DEM is done through the particle-fluid interaction forces. The advantage of coupled CFD-DEM is that it includes all types of interaction between particle-particle, particle-fluid and particle-wall. Moreover, it also provides detailed information at the particle scale, such as particle trajectories and interaction forces, while reducing computational cost comparable to Molecular Dynamics (MD). It is crucial to interpret the fundamental mechanisms that govern the multiscale and multiphase particulate flow. There are two current coupling schemes: unresolved scheme for the typical problems with small particles and resolved scheme for problems with large particles covering few fluid cells.

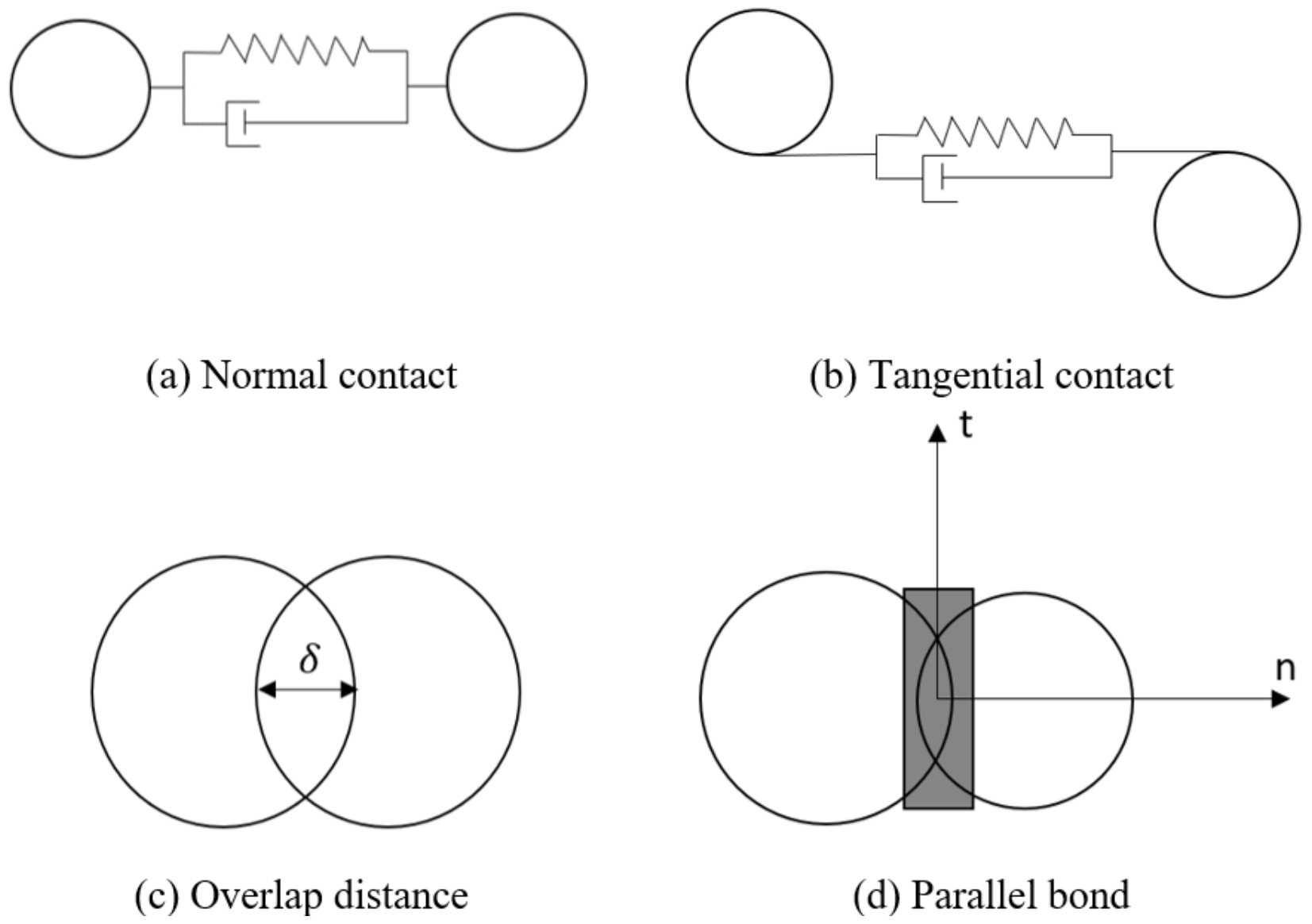

Fig. 1. General contact model with bond.

While the coupled CFD-DEM simulation is suited to simulate the particle-fluid flow, the bonded DEM model is a promising approach to model microstructure of granular, homogeneous and heterogenous materials in problems of fracture or wear. However, most DEM using bond model simulation are quasi-static or dynamics in dry conditions without effect of fluid. There are few research using bond model in the coupled CFD-DEM framework [13], like modelling fibrous materials in the study of clogging in wastewater [14], simulating particle breakage in the jet comminution technology in drilling operations [15], representing the flexible barriers in the investigation of its failure under debris flow [16], and modelling of collapse of buildings in Hong Kong due to the impact of debris flow [17]. Nonetheless, those authors used DEM bond model to represent small particles, a thin barrier or a lattice-ordered structure without examining the accuracy of the fluid flow response. In their studies, the discrete objects may not significantly influence the fluid flow due to their thin and small shapes. In this context, a comprehensive investigation into the fluid flow response over arbitrary shapes of discrete solid objects is lacking. This research will focus on addressing this specific and important question.

The advantage of the proposed approach is to take full advantage of DEM for material modelling and coupled CFD-DEM for particle-fluid simulation. Consequently, it seamlessly simulates the fluid structure interaction (FSI) problem, including the prediction of fracture and damage. The proposed method consists of the following steps:

- A number of discrete particles are generated to fill in an arbitrary geometry of a solid object as shown in Fig. 2. Subsequently, the bonds are added between neighbouring particles. This object is now referred to as a discrete object or discrete domain. The artificial void volume between discrete particles is minimized using the Random Close Packing (RCP) algorithm, which is available in the "granoo-cooker" utility tool provided by the free GranOO workbench [10].
- Coupled CFD-DEM by CFDEM software platform [13] (OpenFOAM [18] -LIGGGHTS [19]) is used for simulating the particulate flow.
- In coupled CFD-DEM simulation, we replace the continuous solid object by a corresponding discrete object. A bond network is added as shown in Fig. 3.

The most important physical features of the problem are:

- the mechanical properties of materials at the macro-scale are reproduced by the micro-scale properties of particles and bonds. Thus, the deformation of a solid object at the macro-scale is represented by the movement of the assembly of the bonded particles at the microscale.
- the interaction between fluid and structure at the macro-scale is the summation of the interaction between fluid and all particles at the micro-scale.

This work is the first attempt to examine the large-scale discrete structure in a fluid environment. We primarily concentrate on examining the fluid response in a general scenario. Afterward, we demonstrate that substituting a continuous solid object with a corresponding discrete one can replicate similar flow behaviour across the entire computational domain. Because the artificial voids exist in the discrete object, a small volume of fluid is trapped in the discrete domain. Indeed, it has a minor influence on the fluid flow in the near wall flow. This last aspect will be further discussed in this paper.

This paper starts by a brief introduction of the numerical approaches including two main parts: Discrete Element Method (DEM) simulating the continuous medium in section 2 and Computational Fluid Dynamics (CFD) in Section 3. The coupled CFD-DEM strategies with resolved and unresolved schemes is summarized in Section 4. Section 5 presents the related benchmarks and discussions for the proposed approaches. Four benchmarks are designed and solved by two different approaches with the same boundary and initial conditions: coupled CFD-DEM with discrete object and traditional CFD approach with continuous object. The final section summarizes the main objectives and conclusion on the proposed approaches.

## 2. Discrete Element Method using bond model to simulate the continuous structure

Discrete element method (DEM) was proposed by Cundall and Strack [2] to simulate granular media like rocks, soil, powder and sands. In DEM, the particle trajectories are tracked in a Lagrangian reference frame by solving Newton's equation (1) for both translational and rotational kinematics. DEM is developed based on the following assumptions:

- particles are considered as pseudo rigid. The particle deformation is represented by the overlap distance between two particles which must be small compared to the particle's radius,
- mass and inertia of materials are concentrated at particle centers, while bonds are massless,
- time step must be chosen smaller than a critical value to ensure stability of the numerical scheme for ensuring that a disturbance can only propagate to the immediate neighbouring elements.

$$m_i \frac{d\mathbf{v}_i}{dt} = \mathbf{F}_{ij} = \sum_j \mathbf{F}_{ij}^C + \sum_k \mathbf{F}_{ik}^{nC} + \mathbf{F}_i^f + \mathbf{F}_i^g$$

$$I_i \frac{d\boldsymbol{\omega}_i}{dt} = \sum_j \mathbf{M}_{ij} \tag{1}$$

where:

- $\mathbf{F}_{ij}, \mathbf{M}_{ij}$ are the interaction forces and torques of particle j acting on particle i.
- $\mathbf{v}_i$ and $\boldsymbol{\omega}_i$ are translational and rotational velocity.
- $\mathbf{F}_{ij}^C$ are contact forces.
- $\mathbf{F}_{ik}^{nC}$ are non-contact forces of particle k acting on particle i, such as van der Waals, electrostatic and magnetic force.
- $\mathbf{F}_i^f$ is the fluid-particle interaction force in coupling problem.
- $\mathbf{F}_i^g$ is the gravitational force.

The interaction between particles is described by interaction force models which can result from a bond (distant interaction) and/or a contact. The bonded forces are the reaction of the bond resulted from the relative movement of particles to their paired ones. The unbonded contact forces are generated from the collision between particles. There are various contact/bond force-displacement models. The advantage of DEM using bond model is that the continuum mechanics laws can be applied on the bond including elasticity, plasticity and fracture behaviour. In the framework of coupling CFD-DEM, all of the interactions between particle-fluid, particle-particle and particle-geometry are included in the contact force term in the equation (1). The particle is always assumed to be in rigid spherical shape throughout this work. A non-spherical or large object is modelled as an agglomeration of smaller spherical ones.

### 2.1. Bond models: flexible fiber

Potyondy and Cundall [3] introduced the Bonded Particle Model (BPM) to include the bond connecting grains inside the rock structure. The fracture of rocks at macroscopic scale is explicitly represented by broking these bonds at the microscopic scale. The BPM has reproduced many features of material behaviour at macroscopic scale including elasticity, fracturing and damage accumulation. Guo, et al. [20] have improved BPM by adding a new damping force to simulate the flexible rod-like fibers. The flexible fibers model has been used by Schramm, et al. [21] to study the dynamics of wheat straw. The more advanced bond types are developed, the more materials and mechanical phenomena DEM is able to model. For rocks, the fracture due to cement bond is replicated. For perfectly elastic materials like silica glass, the tension, torsion, Brazilian and indentation tests have produced very promising result [4-6, 22, 23]. For ductile materials, the preliminary results from tension, compression and buckling test are in good agreement with experiment [9]. The fracture due to thermal mismatch has also been researched [6]. The flexible fiber bond model is chosen throughout this study because its compatibility with the coupled CFD-DEM framework (CFDEM) has been proven and validated [21, 24]. Although the cohesive beam [4-7, 9, 12, 22, 23, 25-27] is more advanced to simulate the continua, the implementation and compatibility with CFDEM framework needs to be investigated further in the future.

The flexible bond model chosen in this study is similar to reference [21] which is slightly different to the original work in reference [20] by using factor of 2 instead of $\sqrt{2}$. Its mathematical model is summarized in the equation (2)-(10):

$$\delta F'^{b}_{n,i} = K_n A_b v_n \Delta t \tag{2}$$

$$\delta F'^{b}_{t,i} = K_t A_b v_n \Delta t \tag{3}$$

$$\delta M'^{b}_{n,i} = K_t I_p \omega_n \Delta t \tag{4}$$

$$\delta M'^{b}_{t,i} = K_n I \omega_t \Delta t \tag{5}$$

$$F^{b}_{n} = \sum_i \delta F'^{b}_{n,i} + 2\beta_{damp}\sqrt{M_e K_n v_n} \tag{6}$$

$$F^{b}_{t} = \sum_i \delta F'^{b}_{t,i} + 2\beta_{damp}\sqrt{M_e K_t v_t} \tag{7}$$

$$M^{b}_{n} = \sum_i \delta M'^{b}_{n,i} + 2\beta_{damp}\sqrt{J_s K_t I_p \omega_n} \tag{8}$$

$$M^{b}_{t} = \sum_i \delta M'^{b}_{t,i} + 2\beta_{damp}\sqrt{J_s K_n I \omega_t} \tag{9}$$

$$K_n = \frac{Y}{l_b}, K_t = \frac{K_n}{2(1-\nu)} \tag{10}$$

where:

- $F^{b}_{n}, F^{b}_{t}$ are the normal and tangential bond forces,
- $M^{b}_{n}, M^{b}_{t}$ are the normal and tangential bond moments,
- $\delta F'^{b}_{n,i}, \delta F'^{b}_{t,i}$ are the normal and tangential incremental bond forces,
- $\delta M'^{b}_{n,i}, \delta M'^{b}_{t,i}$ are the normal and tangential incremental bond moments,
- $K_n, K_t$ are the normal and tangential bond stiffness constants,
- $A_b, l_b$ are the bond cross section area and bond length,
- $M_e, J_s$ are the mass and moment of inertial of individual particles,
- $I, I_p$ are the second area moment and polar area moments of inertia,
- $Y$ is the Young's modulus,
- $\nu$ is the Poison's ratio,
- $v_n, v_t, \omega_n, \omega_t$ are the normal and tangential of relative velocity and angular velocity between the two particles,
- $\beta_{damp}$ is the damping factor.

## 2.2. Discrete element domain

In this work, the discrete domain is generated by the "Random Close Packing" (RCP) method from the "granoo-cooker" utility of the free GranOO workbench [10]. The details of the algorithm is described by André, et al. [4]. The discrete domain must ensure the homogeneity and isotropy of the materials while minimizing the artificial voids between particles. The radius of the discrete elements in the domain complies with a uniform random distribution with a range value of 25% around the mean radius. This value prevents the particle packing of exhibiting ordered arrangements that lead to geometrical anisotropy. The influence of this value on the fracture mode of ductile materials is mentioned in reference [9]. Furthermore, the random arrangement of discrete particles provides a realistic crack path than the ordered lattice [4, 28-30], hence it is more suitable for the materials failure investigation. The "granoo-cooker" produces a compact domain with a void volume fraction around 0.636. At the end, a bond network is added to connect the neighbouring particles. Bond network is characterized by a cardinality number, which can be understood as the average number of bonds per particle. In "granoo-cooker", there are two options to create bond network with cardinality number 6.2 (normal packing that corresponds to a random close packing) or around 13 (corresponding to a Delaunay triangulation algorithm). André, et al. [5] have used Delaunay triangulation with virial stress criteria to produce the Hertzian cone crack in the indentation of silica glass. It is worth to mention that a various way to generate bond type and number of bonds can be found in literature [4, 5, 27, 28, 31, 32]. In this study, the most popular bond network is used with cardinality number of 6.2 and a uniform distribution of radius of 25%.

In the fluid mechanics problem, while the object with a continuous boundary in Fig. 2**a** is used to represent the structure, the discrete domains in Fig. 2**b, c** are likely porous medium where a minor fluid streams can go through the space between particles. Nevertheless, the benchmarks in Section 5 prove that the random arrangement of discrete elements by RCP together with coupled CFD-DEM scheme can minimize the effect of those minor flows and satisfy the fluid flow behaviour as in the case of pure rigid single object. For example, the fluid flow must be static within the zone of discrete domain to ensure rigid object properties. In other words, the fluid velocity should be approaching zero inside the discrete bonded objects.

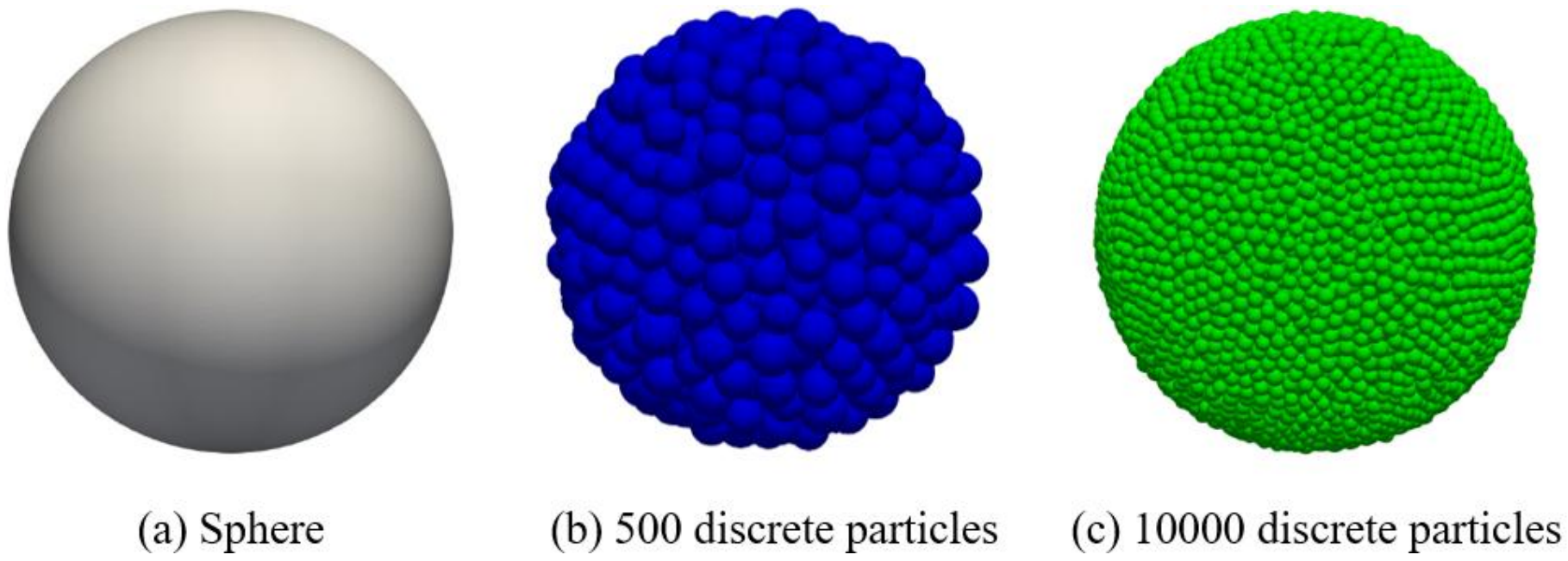

Fig. 2 Decomposition of a sphere into a cluster of discrete elements.

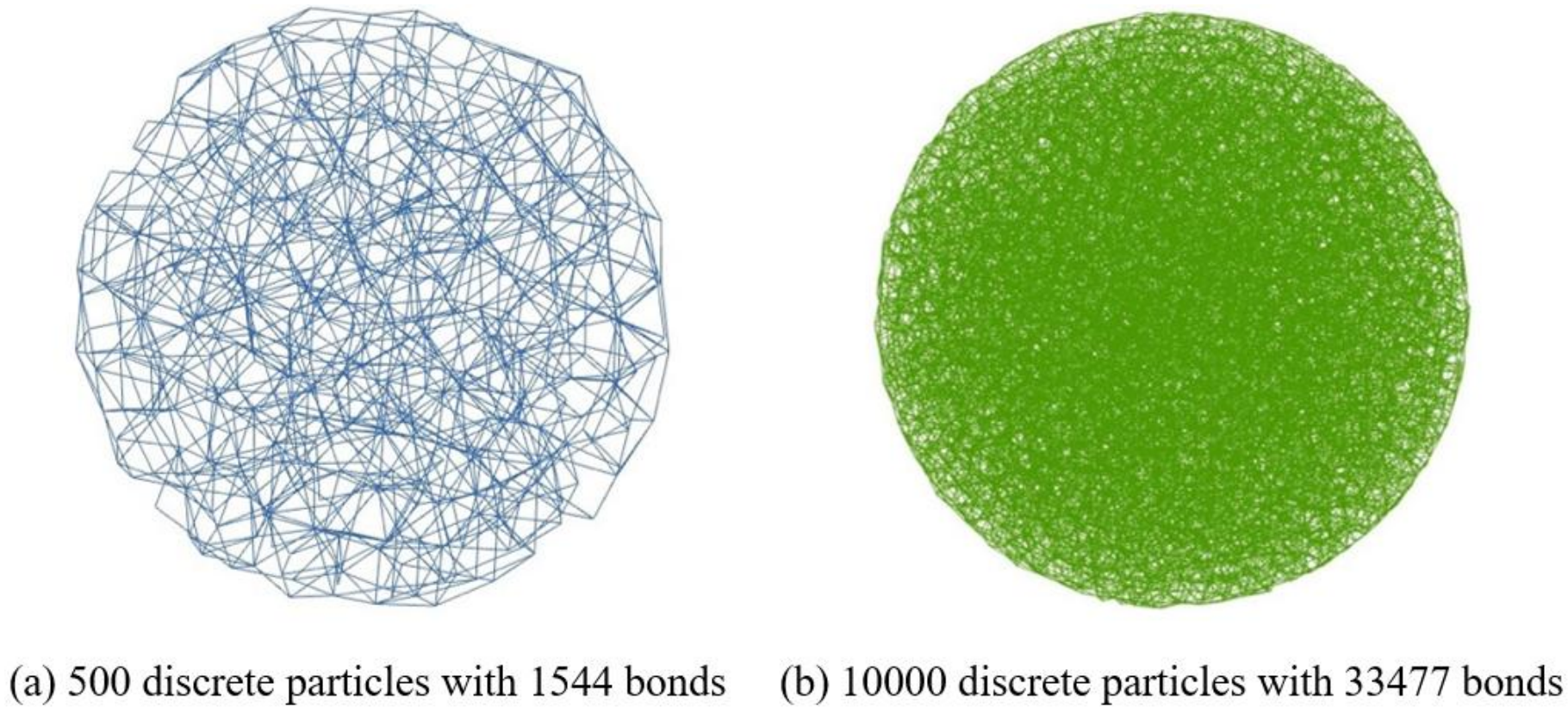

Fig. 3 Bond network in spherical discrete domain.

## 3. Computational Fluid Dynamics (CFD)

### Governing equations

The governing equations for fluid flow, excluding the heat transfer and chemical reactions, are the conservation of mass (continuity equation) and momentum (Navier-Stokes equation) [33, 34]:

Continuity equation:

$$\frac{\partial \rho}{\partial t} + \frac{\partial}{\partial x_i}(\rho\, u_i) = 0 \tag{11}$$

Momentum equation:

$$\rho\left(\frac{\partial u}{\partial t} + u_i \frac{\partial u_j}{\partial t}\right) = -\frac{\partial p}{\partial x_i} + \mu \frac{\partial^2 u_i}{\partial x_i\, \partial x_j} + \left(\frac{1}{3}\mu + \mu_t\right)\frac{\partial}{\partial x_i}\left(\frac{\partial u_i}{\partial x_j}\right) + f_i \tag{12}$$

### Turbulence modelling

In practical life, most fluid flows are turbulent. Turbulence is characterized by several remarkable properties, such as three-dimensionality, unsteadiness, random fluctuations (or irregularity), inclusion of vortices, rotational movement, energy dissipation, and a wide range of length and time scales. There are different turbulence models, including Reynolds-averaged Navier-Stokes (RANS), Large eddy simulation (LES) or Detached eddy simulation (DES). RANS is particularly well-suited for large-scale industrial applications, hence it is chosen for use throughout this study. Among RANS models, k-omega shear stress transport ($k - \omega$ SST) turbulence model, which is a combination of k-omega and k-epsilon models, is favored due to its capability to resolve fluid flow in both the inner and outer region of the boundary layer for a wide range of Reynolds number [35]. This makes it more suitable to simulate turbulent flows in complex geometries and for predicting separation and reattachment of the flow. The application of other models, such as LES, is straightforward by adjusting the settings in fluid solver while keeping the particle properties similar to the corresponding RANS case. The widely recognized Pressure-Implicit with Splitting of Operators algorithm (PISO) [34] is employed to solve the aforementioned equations.

## 4. Coupled CFD-DEM

The particle-fluid flow has greatly gained research interest since they have been observed widely in industry. The combination of computational fluid dynamics (CFD) and discrete element method (DEM) provides a general approach to understand the fundamentals of this multiphase and multiscale problem. Fluid flow is the continuum phase at macroscopic scale described by the Navier-Stokes equations in Eulerian coordinates, while particles are the discrete phase at microscopic scale whose movement are tracked in Lagrangian coordinates. The interaction between particles and surrounding fluid is expressed in the interaction force that includes mainly pressure gradient, drag force and viscous force. Kloss, et al. [36] has introduced the CFDEM framework, which is the coupling of OpenFOAM for CFD and LIGGGHTS for DEM. Its performance has been validated in a number of publications. Both solvers from OpenFOAM and LIGGGHTS can be run efficiently in parallel. Therefore, CFDEM framework is chosen to perform all of the simulations throughout this study. There are currently two coupling schemes: resolved and unresolved.

The general workflow of the coupled CFD-DEM is presented in Fig. 4. The calculation loop starts with DEM solver. Under the bonded and unbonded contact forces, particles are updated with their new positions and velocities. After a number of DEM steps, DEM loop ends, and the particle information is transferred to the fluid solver. It allows for the calculation of the void fraction due to the presence of particles in the fluid cells and the fluid-particle interaction force. Next is the solution of governing equations in fluid solver, followed by the exchange information from the fluid solver to particle solver to continue DEM loop. The calculation loop continues to the final time step. It is essential to mention that the "cohesive beam" in the "DEM Time loop" is used for a general bond type that connects neighboring particles to constitute a continuous material. In this study, the flexible fiber bond is chosen, however, it can be extended to cohesive Euler or Timoshenko beam in future applications.

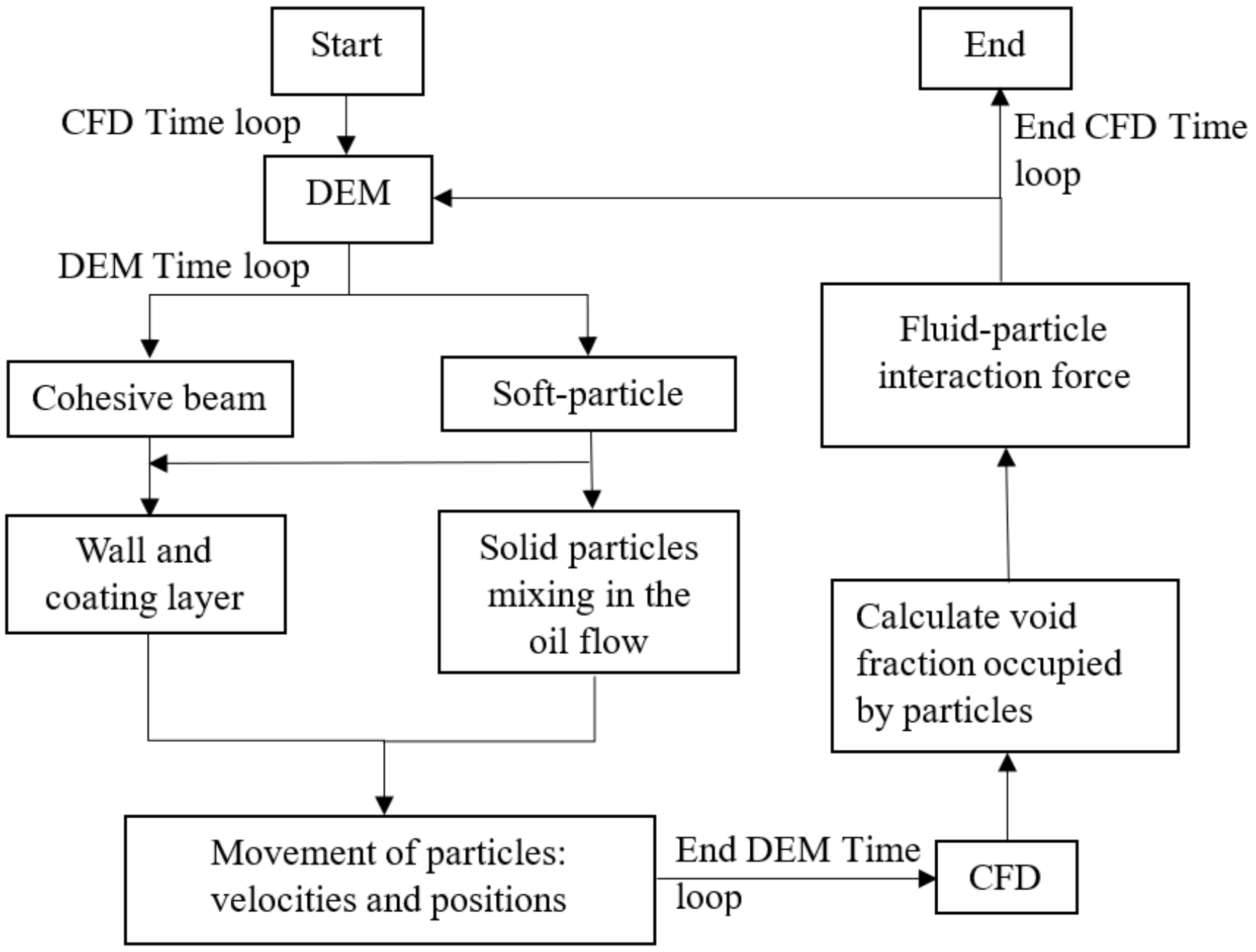

Fig. 4 Calculation loop of coupled CFD-DEM algorithm.

## 4.1. Unresolved scheme

In a general particulate flow, each particle is significantly smaller than a fluid cell depicted in Fig. 5**a**, as a result, the unresolved CFD-DEM scheme is designed for handling the flow containing a large number of particles up to millions. The fluid-particle interaction force is the sum of all types including drag force, pressure gradient force, virtual mass force, Basset force, Saffman force and Magnus force [37]. Among them, the drag, pressure gradient and viscous forces are often considered as the dominant forces in the particle-fluid flow. The other interaction forces have minor contribution and will be excluded from this study. There are three sets of coupling formulations (set I, II and II) and three corresponding models (BFull, A and B) which have been used throughout many research and software. The mathematical models reviewed and classified by Zhou, et al. [38] are summarized in

Appendix C. Based on the review and mathematical formulations of three coupling sets in references [37-39], set II and set I are unconditionally applicable for modelling complex particle-fluid flows. Since model A is corresponds to set II, it has been chosen for all unresolved coupled CFD-DEM simulations in this study.

## 4.2. Resolved scheme

The resolved scheme, which combines the Fictitious Domain Method (FDM) with dynamic local mesh refinement, was implemented in the CFDEM framework by Hager [40] to handle flows with particles larger than the fluid cells (Fig. 5**b**). Dynamic local mesh refinement, as shown in Fig. 7**b** and Fig. 7**c**, was used to refine the mesh around each particle, ensuring that each particle diameter covers at least eight fluid cells. When a particle moves to a new cell, the previously refined cells are unrefined, and the current cells are refined. This process continues along the particle's movement path. At first, Hager [40], [41] used a zero-one representation for void fraction field with only two values, one or zero, to mark the cells whose centers are covered by the body of the large particle. Later, a stair-step method (or smooth particle representation model) with a weighting factor was also introduced by Hager to take into account the degree which a cell is covered. The basic resolved scheme without smoothing function is used throughout this study.

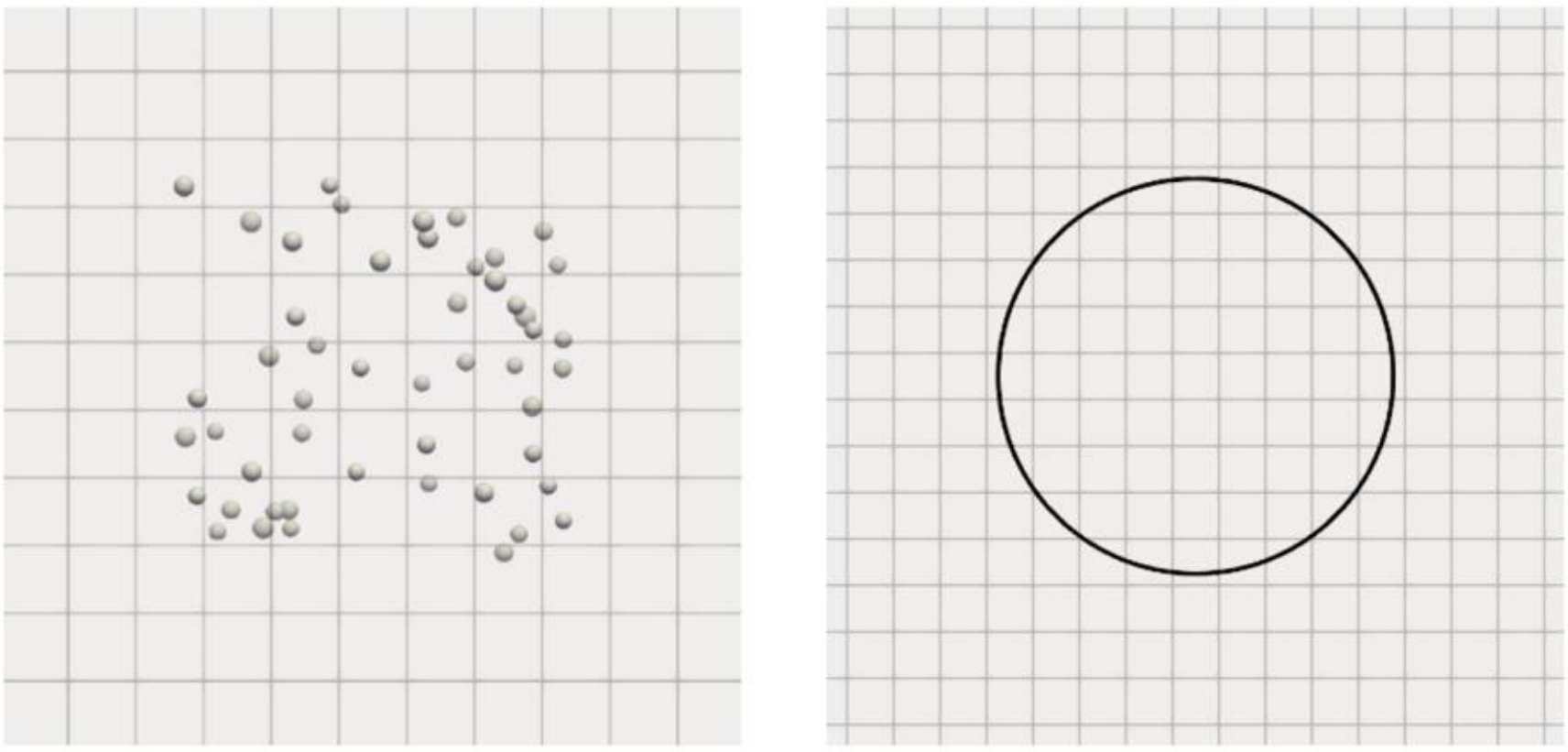

(a) Unresolved CFD-DEM (b) Resolved CFD-DEM

Fig. 5 Demonstration of fluid cell and particle sizes in two coupled CFD-DEM schemes.

## 4.3. Finnie erosion equation

The Finnie model [42, 43] for erosive wear describes the relationship between erosive wear and the rate of kinetic energy of a single particle impact on a surface:

$$E = kU_p^2 f(\gamma) \tag{1}$$

where E is a dimensionless wear rate, k is constant, $U_p$ is the particle impact velocity, $f(\gamma)$ is the function of angle between the incoming particle's trajectory and surface:

$$\begin{aligned} f(\gamma) &= \frac{1}{3}\cos^2(\gamma) && \text{if} \tan(\gamma) > \frac{1}{3} \\ f(\gamma) &= \sin(2\gamma) - 3\sin^2(\gamma) && \text{if} \tan(\gamma) < \frac{1}{3} \end{aligned} \tag{2}$$

The equation above applies for a single particle, hence when it is used in conjunction with a mass flow rate of particles $\dot{m}$, the eroded mass EM in kg during one time step $\Delta t$ is calculated as:

$$EM_{HS} = E\,\dot{m}\,\Delta t \tag{3}$$

## 5. Benchmarks and discussion

Throughout this study, different types of fluid flow over solid object are investigated using two approaches with the same initial and boundary conditions: coupled CFD-DEM with deformable discrete objects and CFD with rigid continuous walls. The results from CFD simulation provide quantitative validation. In a CFD simulation, the solid object is considered rigid with a continuous boundary, so the mesh covers the computational domain excluding the solid volume and must conform to the solid boundary. In a coupled CFD-DEM simulation, the solid object is replaced by an agglomerate of discrete particles. Consequently, the mesh covers the entire computational domain, with particles treated as voids within fluid cells. The difference in mesh between different approaches (CFD, resolved coupling scheme and unresolved coupling scheme) is illustrated in Fig. 7.

The simulation cases are arranged in the following order, starting from a simple to a complex fluid flow: At first, the rigid continuous solid object is replaced by a rigid discrete domain where all particles are fixed. These particles are called "wall-particles" because they are bonded together and fill the volumetric space of the original continuous solid object. This test represents the flow over a rigid and fixed object. A laminar flow over a spherical object and turbulent flow over a non-spherical or blunt object, such as a square cylinder, are introduced. Next, the combination of a continuous wall and a discrete wall is presented. The purpose of this example is to use a discrete wall in areas where we want to focus on the mechanical behaviour of the solid structure, while neglecting parts of the continuous wall where deformation is negligible. After proving that the replacement of rigid continuous object by a rigid discrete one provides fluid flow field in good agreement with the traditional CFD approach, the next case is to allow particles to move. There are now two types of particles: "wall-particles" and "free-particles". This model is called "full discrete model" or "full particle-scale model". The "wall-particles" are bonded together to represent the solid object, while the "free-particles" correspond to the stream of sand or dispersed solid particles mixed in the fluid flow. This type of problem involves three-way coupling of Fluid-Structure-Particle Interaction where the movement of "wall-particles" reflects the mechanical properties of solid and its deformation under the impact of fluid flow and "free-particles" stream. This is an application of the proposed full discrete or full particle scale model to study erosion.

The time step in each case must be chosen to satisfy the Courant-Friedrichs-Lewy (CFL) criterion, ensuring that the CFL number is less than 1 to maintain numerical accuracy. It is noted that the gauge pressure, which is relative to ambient pressure, is used in all of fluid flow simulation. The second order discretization scheme is used for gradient, divergence and Laplacian terms in the governing equations to achieve high accuracy. In mechanical problems without thermal effect or chemical reaction, the validation focuses on the distribution of velocity and pressure, particularly in the area near solid object, as well as streamlines of the flow over the continuous and discrete one. The fluid properties are listed in Table 5.

### 5.1. Laminar flow over sphere

In this section, laminar flow over a stationary spherical solid object is investigated. This can serve as an initial validation and provides preliminary parameter estimates for the full particle scale model. As mentioned in the reference [40], the laminar flow with Re=100 passes through the sphere without any instabilities and results in a symmetric wake downstream of the body. In general, the computational domain for external flow must be large to ensure that disturbances from far-field boundaries do not affect the near-wall flow around the object. The computational domain is shown in Fig. 6, and the related sizes and lengths are listed in Table 6. In this example, the front and side lengths (L1, W1, W2, H1 and H2) have been chosen to be ten times the sphere diameter, while the wake size (L2) is twenty times the sphere diameter. Both resolved and unresolved coupling schemes will be examined. Three coupled CFD-DEM cases have been set up together with one CFD simulation as a benchmark:

- **Case 1:** resolved CFD-DEM scheme with only one large spherical particle with size $d = 0.02$ m $= 20$ mm.
- **Case 2:** resolved CFD-DEM scheme with a large number of spherical particles that compose the original spherical volume. Since the fluid flow around individual particles is resolved in this scheme, the number of discrete particles will affect the reproduction of streamlines as well as the flow field, including velocity and pressure. Four different simulation cases are set up to investigate the influence of the number of discrete particles:
    - (2a) with 500 particles having average size (diameter) $d_{avg}$~2.3 mm
    - (2b) with 1,000 particles having average size (diameter) $d_{avg}$~1.6 mm
    - (2c) with 5,000 particles having average size (diameter) $d_{avg}$~1 mm
    - (2d) with 10,000 particles having average size (diameter) $d_{avg}$~0.8 mm
- **Case 3:** unresolved CFD-DEM scheme with 50,000 small spherical particles having an average size $d_{avg}$~0.5 mm. This coupled scheme requires the particle size to be smaller than the fluid cell, hence, the number of particles is much greater compared to the resolved coupled scheme.

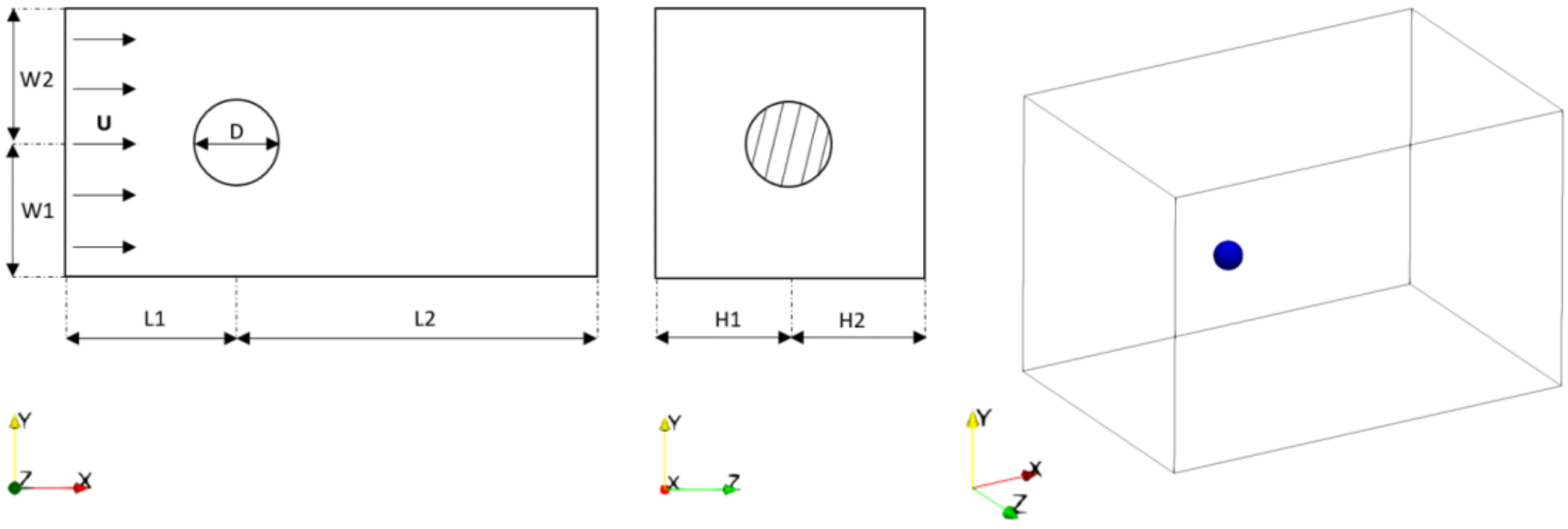

(a) Oxy plane (b) Oyz plane (c) 3D view

Fig. 6 Computational domain of flow over sphere.

The meshes of four simulations are shown in Fig. 7. In CFD simulations, the mesh needs to conform to the boundary of the solid object, while in coupled CFD-DEM simulations, the mesh covers the entire computational domain and overlaps the solid object. In the resolved case with a single large particle, the mesh is mainly refined around the surface boundary of that particle. In the resolved case with many neighboring particles, almost all mesh cells covering the particles are refined. Conversely, the mesh in the unresolved scheme is very simple without any local refinement or topological match requirement.

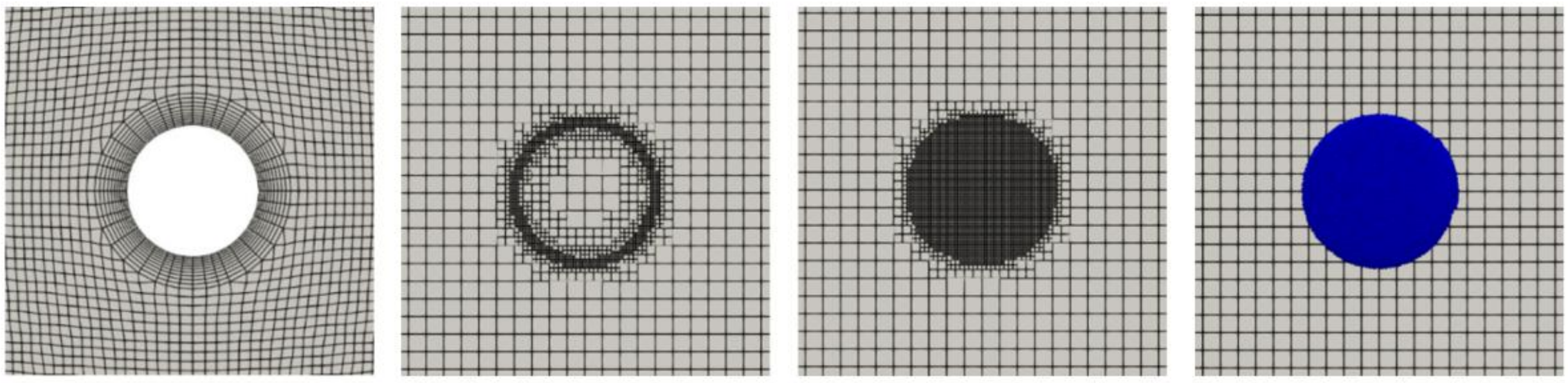

(a) CFD

(b) resolved scheme with only 1 particle (case 1)

(c) resolved scheme with 5,000 particles (case 2c)

(d) unresolved scheme (case 3)

Fig. 7 Mesh types in each simulation.

The snapshot of fluid flow streamlines on a middle cross-section at time t=200s are shown in Fig. 8. The velocity and pressure distribution across the sphere are measured along line AB, where the y-coordinate varies from -0.01 to 0.01 m. Line CD, which represents the wake length, are marked in Fig. 8a for post-processing and comparison purpose. In all simulations, a symmetric pair of vortices is observed in the wake behind the sphere. The velocity magnitude in the whole computational domain ranges from 0 to 5.6e-3 m/s. The fluid flow impacts normally to the sphere, hence the stagnation zone with lowest velocity or highest pressure is observed in front of the sphere. When the fluid flows around the sphere, its velocity is increased, so we can see the highest velocity and lowest pressure on the side of the sphere, which is the top and bottom part from the viewpoint. The length of wake in all cases and the difference in percentage compared with the CFD simulation are summarized in Table 1. It is clearly seen that the streamline profiles generated by the resolved scheme with only 1 large particle, 5000 and 10000 particles most closely match the CFD benchmark with a relative difference in wake length 26%, 32 % and 35%, respectively. In contrast, the unresolved scheme significantly overpredicts the wake length that is almost double the CFD result. It should be noted that although the relative percentage is quite high, the absolute value of deviation is very small and negligible in real life.

In the resolved CFD-DEM model, the simulation with a single large particle (case 1) results in smooth streamlines. In case 2a with 500 particles filled in the volume of the sphere, the streamlines near the boundary of the sphere exhibit slight disturbances. Because the summation of 500 particles does not completely fit with the spherical volume, they leave some void resulting in fluid flow penetration. As a result, the streamline of fluid flow is not as smooth as CFD benchmark. When the number of particles is increased to 5000 and 10000, the original volume seems to be better filled with almost smooth boundary and very little void left. Consequently, the streamline becomes smoother. In the unresolved scheme, because the flow around each particle is not resolved explicitly, the streamline does not match perfectly around the boundary of the sphere. Since the near wall flow is unresolved, the flow passes through the artificial space between neighboring particles. The influence of boundary wall in CFD is implicitly included by fluid-particle interaction forces in coupled CFD-DEM method. Although the wake behind the sphere in case 3 is longer than the other cases, the unresolved coupled CFD-DEM simulation gives a quite good prediction for the average velocity and pressure in the whole computational domain with little disturbance in the near wall region.

Table 1 Length of wake behind the sphere demonstrated as line CD.

| Case | Coupled CFD-DEM scheme | Number of particles | Length of wake (m) | Absolute difference from CFD (m) | Relative difference from CFD (%) |
|---|---|---|---|---|---|
| CFD | | | 0.0111 | | |
| Case 1 | Resolved | 1 | 0.014 | 0.0029 | 26.13 |
| Case 2a | Resolved | 500 | 0.0169 | 0.0058 | 52.25 |
| Case 2b | Resolved | 1,000 | 0.0154 | 0.0043 | 38.74 |
| Case 2c | Resolved | 5,000 | 0.015 | 0.0039 | 35.14 |
| Case 2d | Resolved | 10,000 | 0.0147 | 0.0036 | 32.43 |
| Case 3 | Unresolved | 50,000 | 0.0215 | 0.0104 | 93.69 |

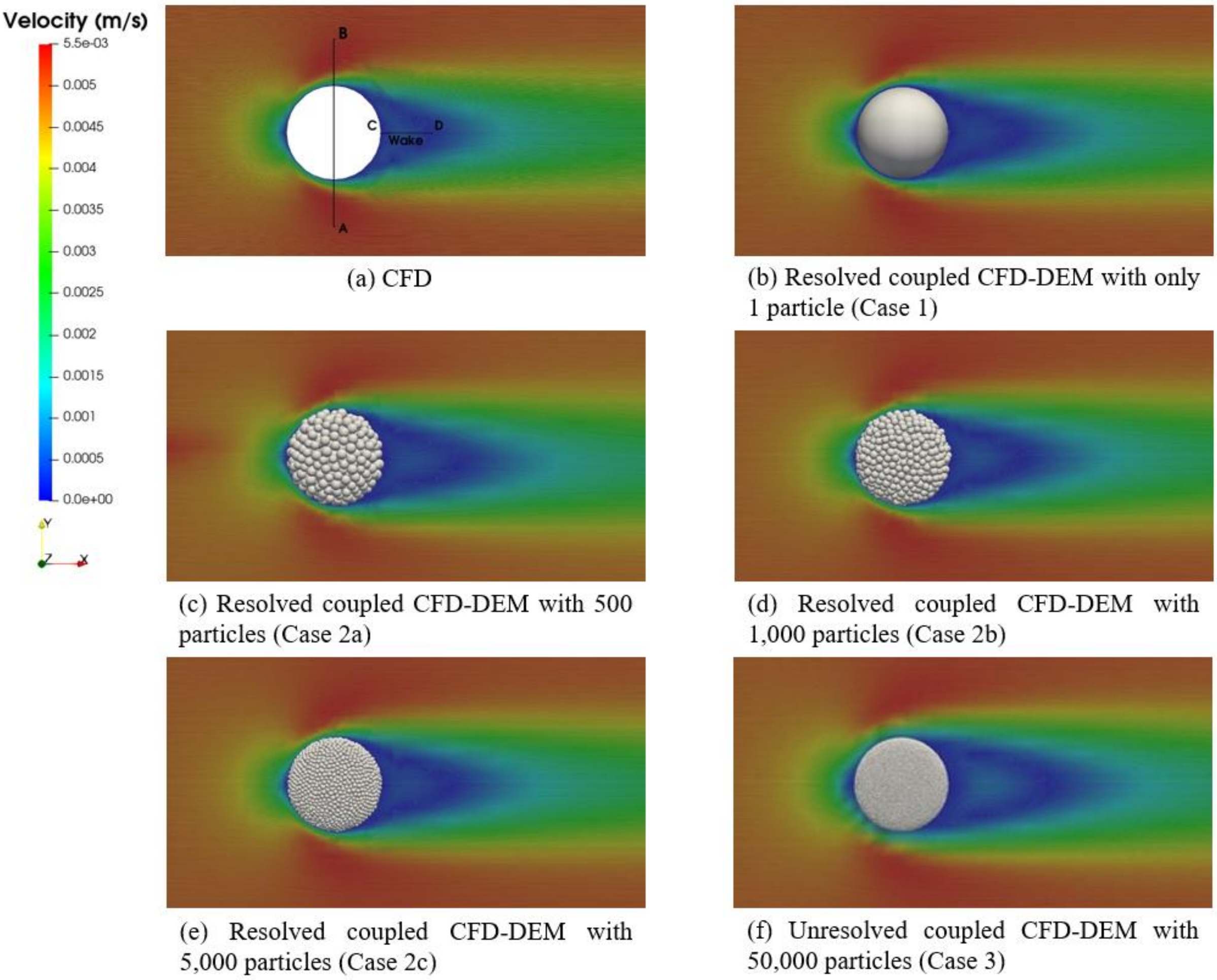

(a) CFD

(b) Resolved coupled CFD-DEM with only 1 particle (Case 1)

(c) Resolved coupled CFD-DEM with 500 particles (Case 2a)

(d) Resolved coupled CFD-DEM with 1,000 particles (Case 2b)

(e) Resolved coupled CFD-DEM with 5,000 particles (Case 2c)

(f) Unresolved coupled CFD-DEM with 50,000 particles (Case 3)

Fig. 8 Streamlines reproduction colored by velocity magnitude at t=200s.

In order to perform quantitative comparisons, the velocity and gauge pressure distributions along line AB, which crosses the sphere's surface, are presented in Fig. 9 and Fig. 10, respectively. In the CFD simulation, fluid does not penetrate the sphere's region, resulting in no measured values for velocity or pressure within the sphere. However, in the coupled CFD-DEM simulation, the mesh covers the full computational domain, and the particles fill the sphere region, allowing for velocity and pressure values to be obtained in those locations. It can be seen from Fig. 9, the resolved scheme predicts the velocity distribution very closed to the CFD benchmarks, while the unresolved scheme results in similar profile but with small discrepancies. As the number of particles increases, the velocity profile resulted from the resolved CFD-DEM case moves closer to the CFD benchmark and the resolved case with one large particle. The unresolved CFD-DEM appears to overpredict the velocity in the stagnation zone because it does not analyze the flow in the space around individual particles. In Fig. 10, inside the sphere region in coupled CFD-DEM simulation, the pressure values do exist. It can be seen from Fig. 10**a**, the case 2a, which is the resolved scheme with 500 particles, results in a pressure larger than the others. Fig. 10**b** shows a closer look of the pressure profile across the sphere region of all other cases excluding case 2a due to the significant difference. The reason for the significant difference in gauge pressure values is that the figure is zoomed in on a specific value range for comparison. However, the absolute values, which are approximately 2e-4 Pa in case 2a and 2e-5 Pa in the others, are nearly zero and negligible.

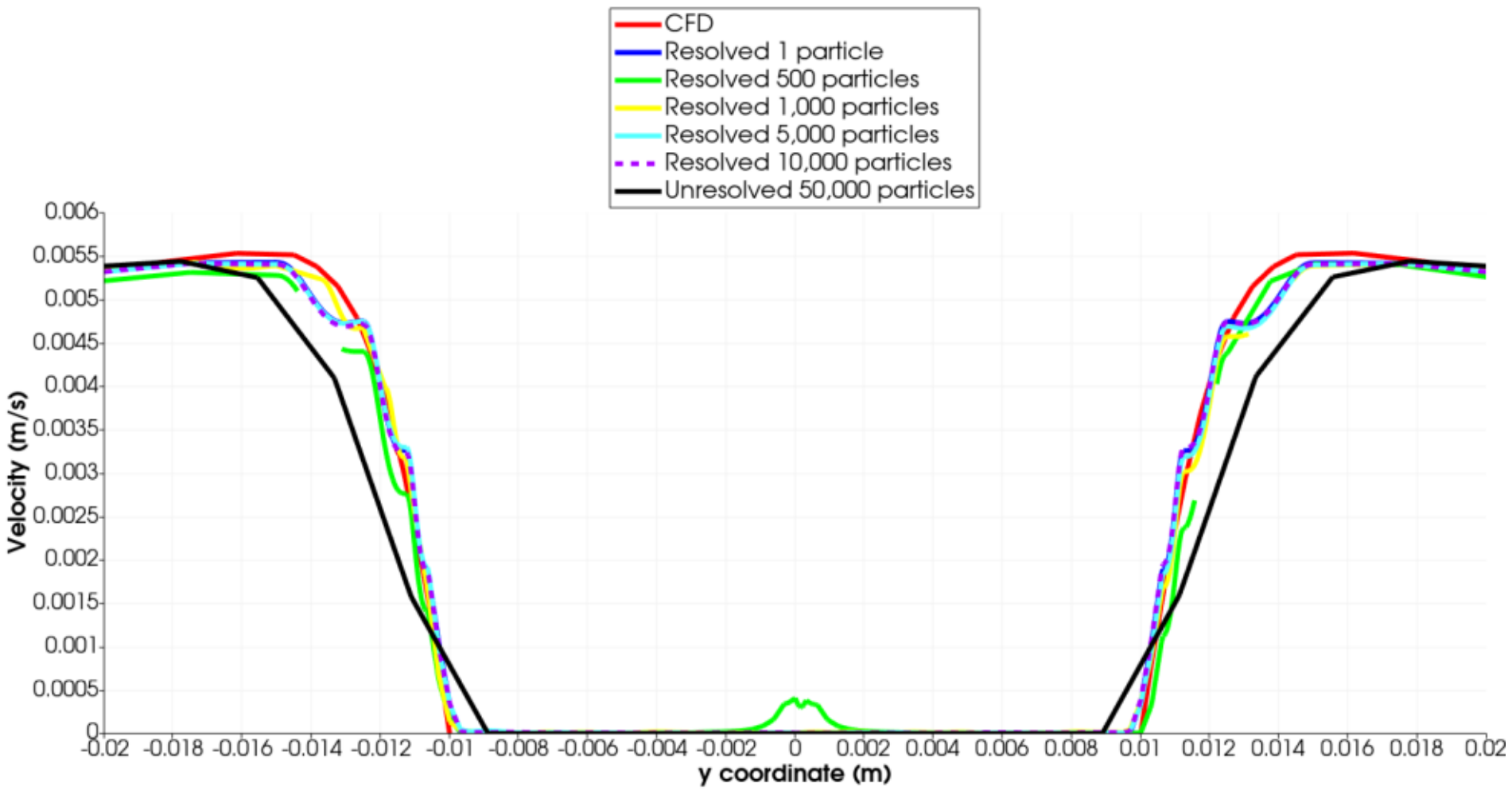

Fig. 9 Velocity magnitude on the line across the sphere surface connecting two points A(0, -0.02, 0) and B(0, 0.02, 0).

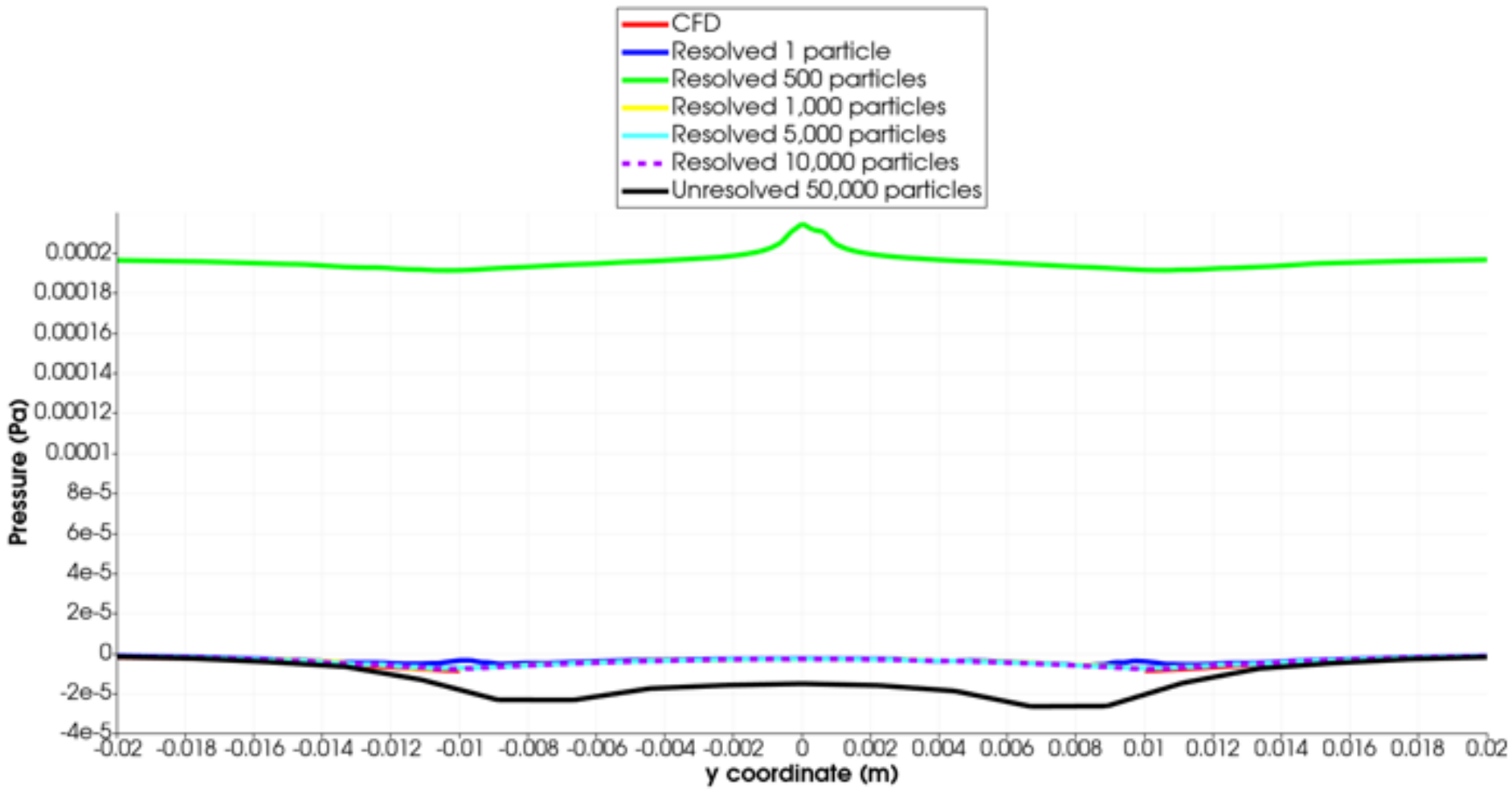

(a) Pressure

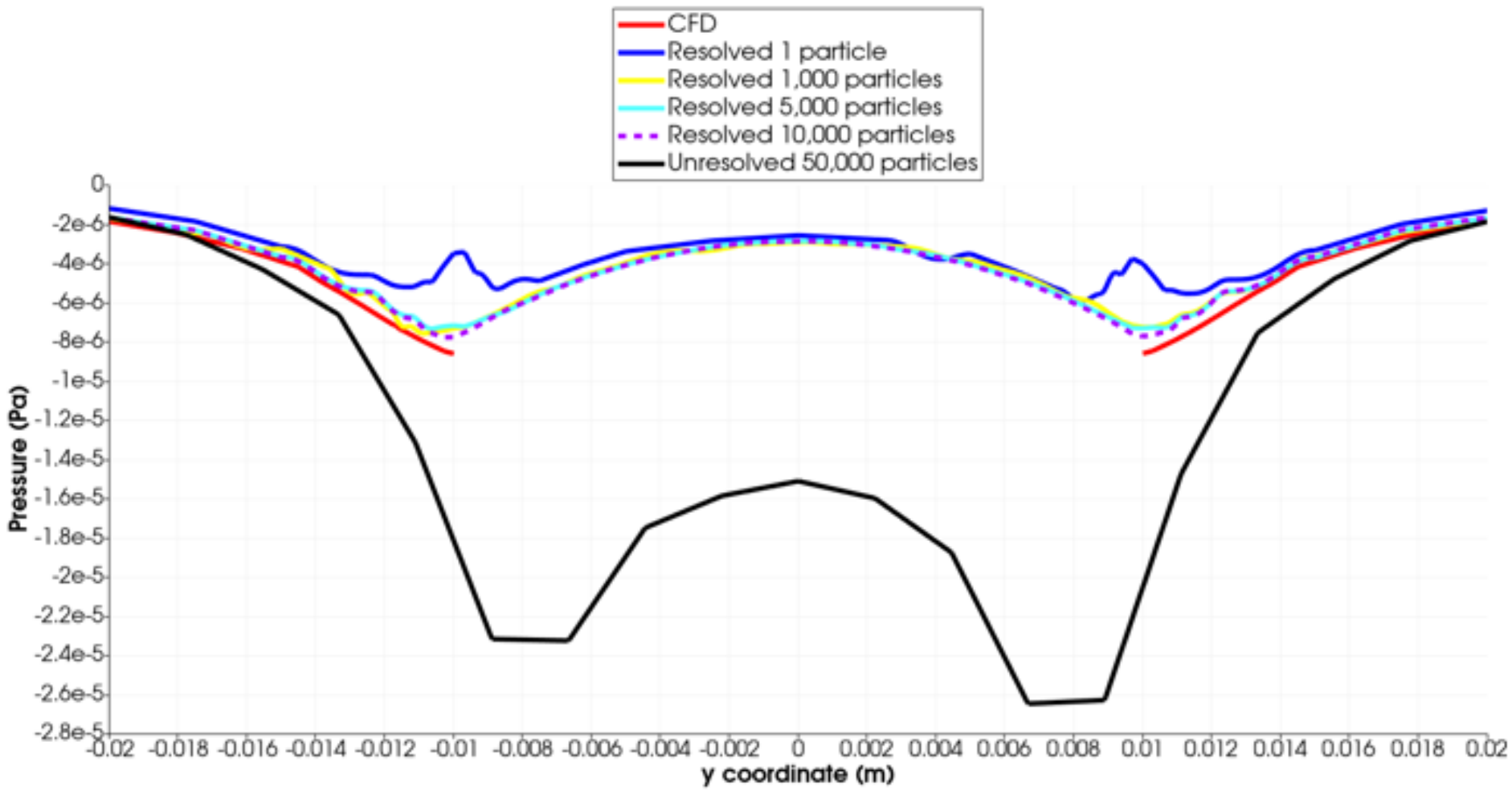

(b) Pressure, excluding case of resolved CFD-DEM with 500 particles

Fig. 10 Gauge pressure on the line across the sphere surface connecting two points A(0, -0.02, 0) and B(0, 0.02, 0).

In terms of the computational time listed in Table 10 , Appendix B, the resolved case requires expensive computational resources, while the unresolved coupled CFD-DEM is much faster and just a little slower than the CFD simulation. The trade-off of not resolving fluid flow around individual particles in the unresolved scheme is the faster computing time. Based on these findings, it can be concluded that the unresolved scheme is more suitable for large-scale industrial problems, whereas the resolved CFD-DEM simulation with requirement of huge computational resources is applicable for fundamental study of mechanisms.

### 5.2. Turbulent flow over blunt object with sharp edges (square cylinder)

This section aims to demonstrate the effectiveness of the full particle scale approach in modeling turbulent flow over a square cylinder with blunt body and sharp edges. The computational domain is described in Fig. 11, while the related sizes and lengths are listed in Table 7, Appendix A. This is a dynamic problem with the vortex shedding in the wake behind the square cylinder. Since the unresolved scheme is better suited for industrial applications, its performance is further investigated here. In the laminar flow studied in the previous benchmarks, two simulation results from CFD and coupled CFD-DEM are in good agreement in terms of average velocity and pressure fields. However, in this problem, the more important physics to reproduce is the dynamic behavior of vortex shedding. Fig. 12 and Fig. 13 show the snapshots of vortex street at different time steps from CFD and unresolved CFD-DEM simulation, respectively. From these two figures, it is seen from both simulations that the flow accelerates as it passes over the object, creating a low velocity zone immediately behind it. Subsequently, after a certain period of time, the instability behavior begins to manifest, resulting in the cyclic shedding of vortices.

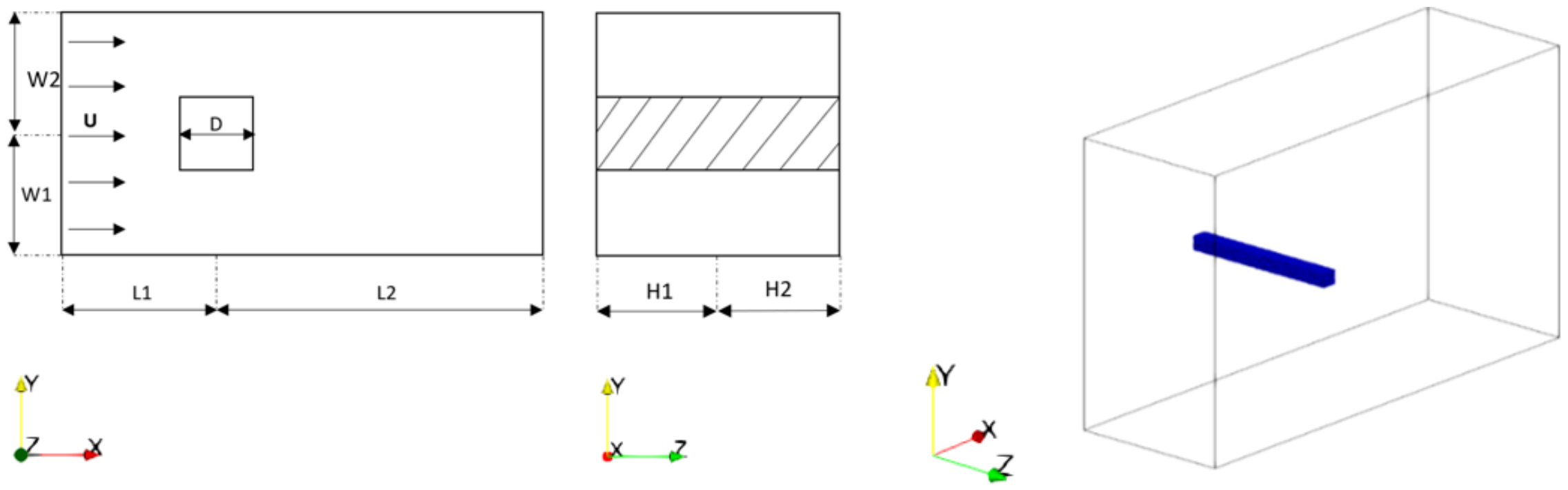

(a) Oxy plane (b) Oyz plane (c) 3D view

Fig. 11 Computational domain of flow over blunt object (square cylinder)

In CFD simulation (Fig. 12), the vortex shedding begins to occur in the wake behind the square cylinder at time step of t=5s, then fully developed quickly since t=10s. The state of the vortex street at t=10s is similar to that at t=25s, as are the states at t=15s and t=30s. Therefore, it can be roughly estimated that the cycle time from the CFD simulation is approximately 15 seconds. As the vortices propagate, the first vortex in the bright zone (indicated by the green color) moves from the top position in Fig. 12**b** to the bottom position in Fig. 12**d** in less than 10 seconds, which is approximately half of the cycle time, as anticipated. The width measured in the vertical direction (y-axis) from the top peak to the bottom peak of vortex street approximate to 0.123 m.

It can be seen clearly in Fig. 13 that although the unresolved coupled CFD-DEM scheme can reproduce dynamic vortex street, it is delayed compared to the CFD simulation. The instabilities occur late since the time step t=35s and appear to be fully developed by t = 40s. The state of velocity contour shown in (Fig. 13b) share similarities with that in (Fig. 13d), so do the states depicted in (Fig. 13c) and (Fig. 13e). Therefore, it can be inferred that the cycle time is approximately 25-35 seconds, nearly twice that of the CFD case, which has an estimated cycle time of 15 seconds. This difference in cycle time is resulted from the difference in velocity distribution in the wake from the unresolved CFD-DEM case. Based on the color bar, the red color indicates the high-speed zone, while the green color indicates the slow-speed zone. The CFD results show that the wake is dominated by high-speed zones, resulting in many separate regions with green color. This leads to faster fluid flow in the wake and the shedding of vortices at low cycle times. In contrast, the velocity contour in the coupled CFD-DEM indicates that the wake is dominated by low-speed zones and smaller fraction of regions in red color. Thus, it results in low fluid flow in the wake and the cycle time of vortex shedding estimated from the coupled CFD-DEM is longer.

It should be emphasized that in the CFD case, the boundary conditions for turbulent quantities such as turbulence kinetic energy (k) and specific turbulent dissipation rate (omega) are applied directly on solid surface boundary through near wall treatment, described in mathematical equations, while it is not explicitly included in coupled CFD-DEM. Their effects on flow field properties are implicitly included by the interaction with the particle phase in the governing equation. Three interaction forces, namely the pressure gradient, viscous and KochHill drag force, are used throughout this study. Therefore, it can be concluded that the unresolved coupling scheme, with the selected interaction forces, may not reproduce accurately the dynamic instability near the wall that causes the delay of the vortex sheet, as observed in Fig. 13.

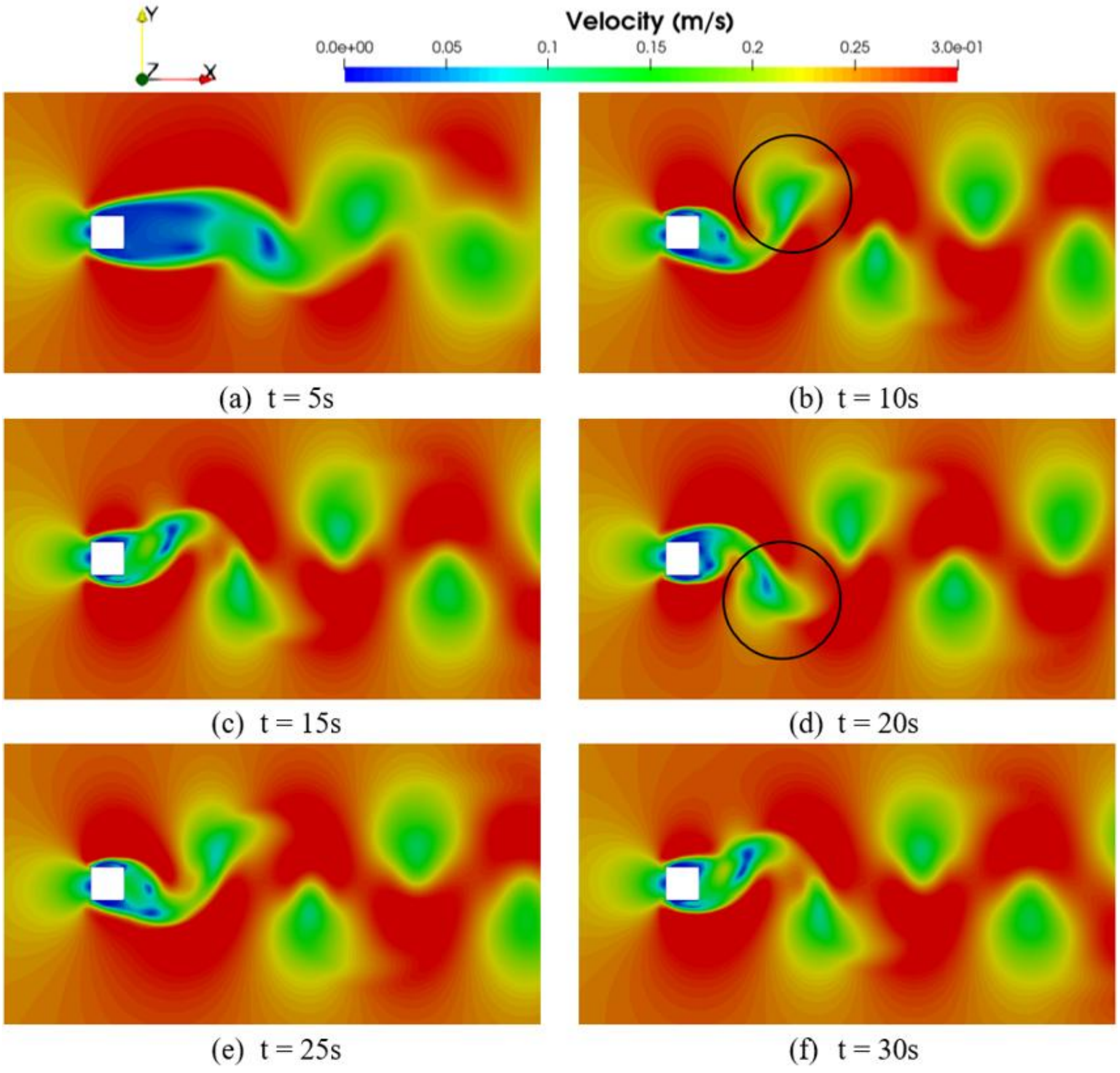

Fig. 12 Vortex shedding of turbulent flow over square cylinder described in velocity contour from CFD simulation. Time step is measured in second (s). The first and nearest vortex from the object is circled in black colour.

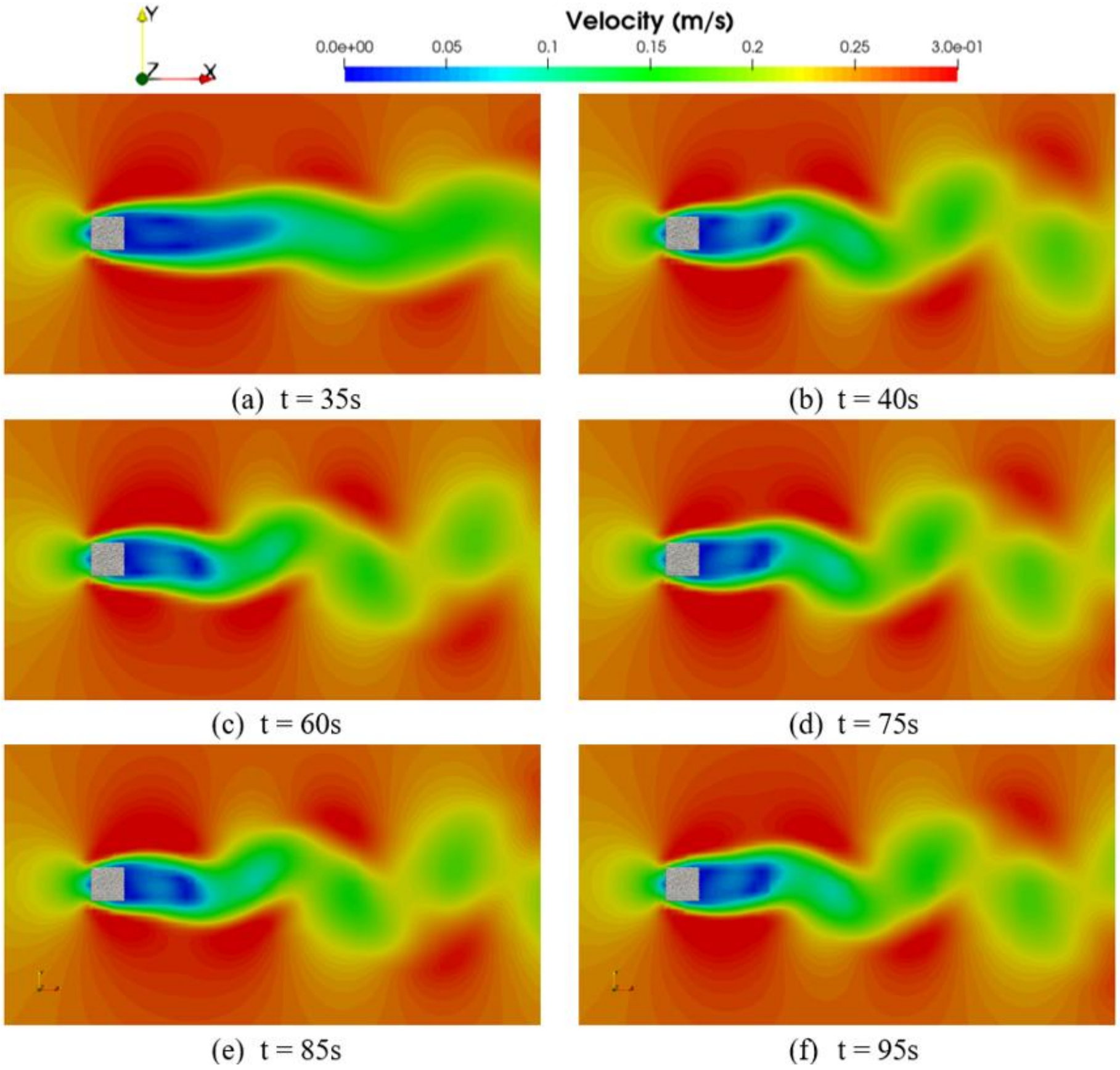

Fig. 13 Vortex shredding of turbulent flow over square cylinder described in velocity contour from coupled CFD-DEM simulation with 100,000 particles. Time step is measured in second (s).

### 5.3. Combination of rigid continuous and deformable discrete wall: Erosion test rig configuration

In two previous examples, different types of fluid flow over spherical and non-spherical discrete objects in both laminar and turbulent mode have been studied and validated by the corresponding CFD simulation. Table 10 in Appendix A summarizes the computational time and resources allocated to all test cases using resolved and unresolved coupled CFD-DEM schemes. While the resolved coupled scheme is more suitable for fundamental studies or problems including a limited number of large particles, the unresolved scheme is computationally feasible and more suitable for large-scale industrial applications. Therefore, the unresolved scheme will be applied to investigate the behaviour of fluid flow in the erosion test rig. In most simulations, it is not necessary to replace all continuous objects with discrete ones, as this would significantly increase the demand for computational resources. Erosion prediction remains a challenging problem due to the complex physics of multiphase and multiscale flows. A general erosion simulation includes three main steps [44]: (1) Flow pattern prediction, (2) Particle tracking and (3) Erosive wear rate and pattern prediction. The combination of continuous and discrete walls in this section serves as the first step in validating the velocity and pressure field in the erosion test rig. The base configuration of the erosive wear test conducted in references [35, 45, 46] is the carrier fluid (water, air, oil) mixed with sand particles from nozzle impacting normally to a plate. In this type of simulation, we focus on the deformation of the plate and erosion occurring on its surface. Therefore, the plate is replaced with a discrete wall, while the nozzle remains a continuous wall. This section simulates water flow from a nozzle impacting a plate. The insertion of free sand particles into the fluid stream will be discussed in section 5.4. The configuration is shown in Fig. 14 with dimension listed in Table 8, Appendix A.

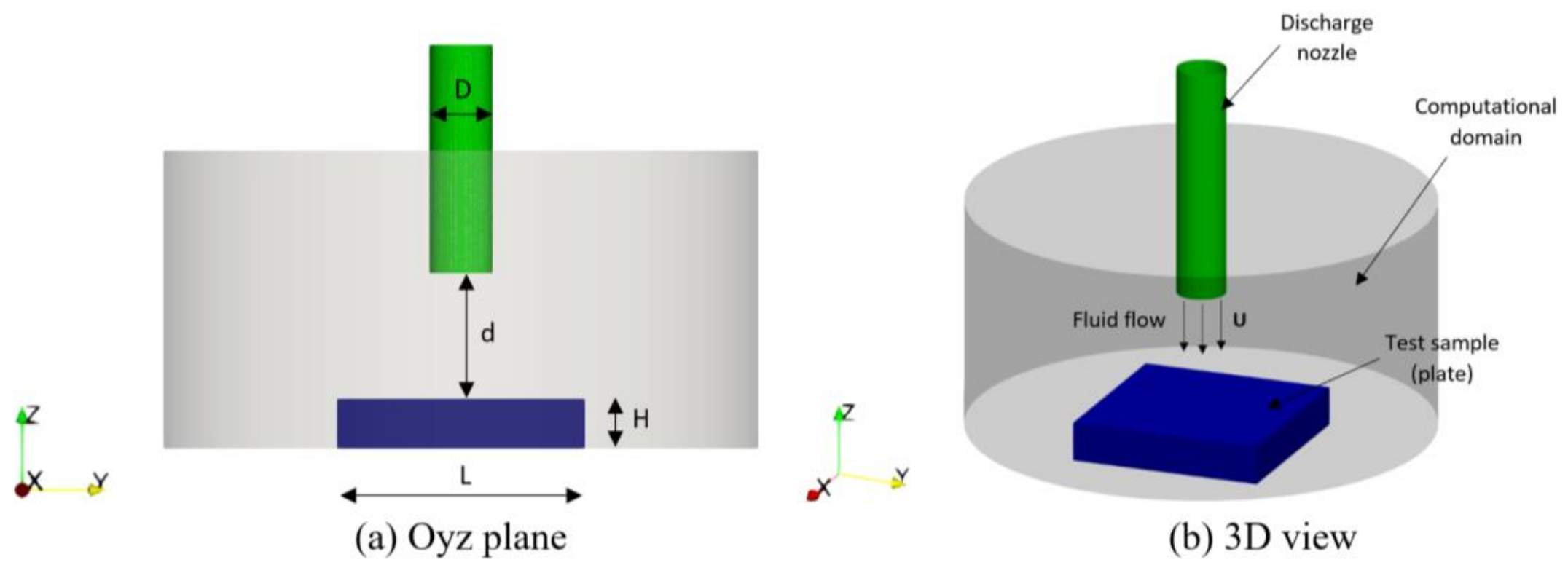

Fig. 14 Configuration of water flow (**U**) from nozzle (green) impacting a plate (blue).

The overall distribution of velocity and pressure in Fig. 15 and Fig. 16 match well with the refence profiles in [35]. As can be observed from Fig. 15 and Fig. 16, the unresolved CFD-DEM and CFD simulation shows similar velocity, pressure contour and streamlines. The stagnation zone with zero fluid velocity and highest pressure is at the intersection of jet flow from nozzle and plate. In both cases, after impacting the plate, the flow tends to spread outwards. In CFD simulation, the highest velocity is recorded at about two times of the nozzle radius from the center of nozzle. Whereas, at that location, the unresolved coupled CFD-DEM shows a smaller value.

Because the CFD simulation applies no slip wall boundary condition to the sample surface, the velocity of fluid flow on the surface is equal to 0. Therefore, it is more accurate to compare the velocity values measured at a small distance (~ 0.5 mm) from the sample surface, illustrated in Fig. 17 and Fig. 18. The pressure distribution in the two simulation cases closely matches, while the unresolved CFD-DEM simulation underpredicts the magnitude of fluid velocity, with a value of 0.8 m/s compared to 1 m/s in the first case, and 3.8 m/s compared to 4.6 m/s in the second case. Although there is a slight difference in value prediction, the velocity profile across the sample surface, which resembles a V-shape, and the location of the highest velocity, which is located around 6 mm from the center, are similar to the pattern of CFD results. The oblique angle configurations (45$^{o}$ and 60$^{o}$) are also simulated under the condition of maintaining a fixed distance from the nozzle to the center of the sample surface. Their velocity and pressure contours shown in Fig. 19 qualitatively match well with the patterns shown in references [35, 45, 46].

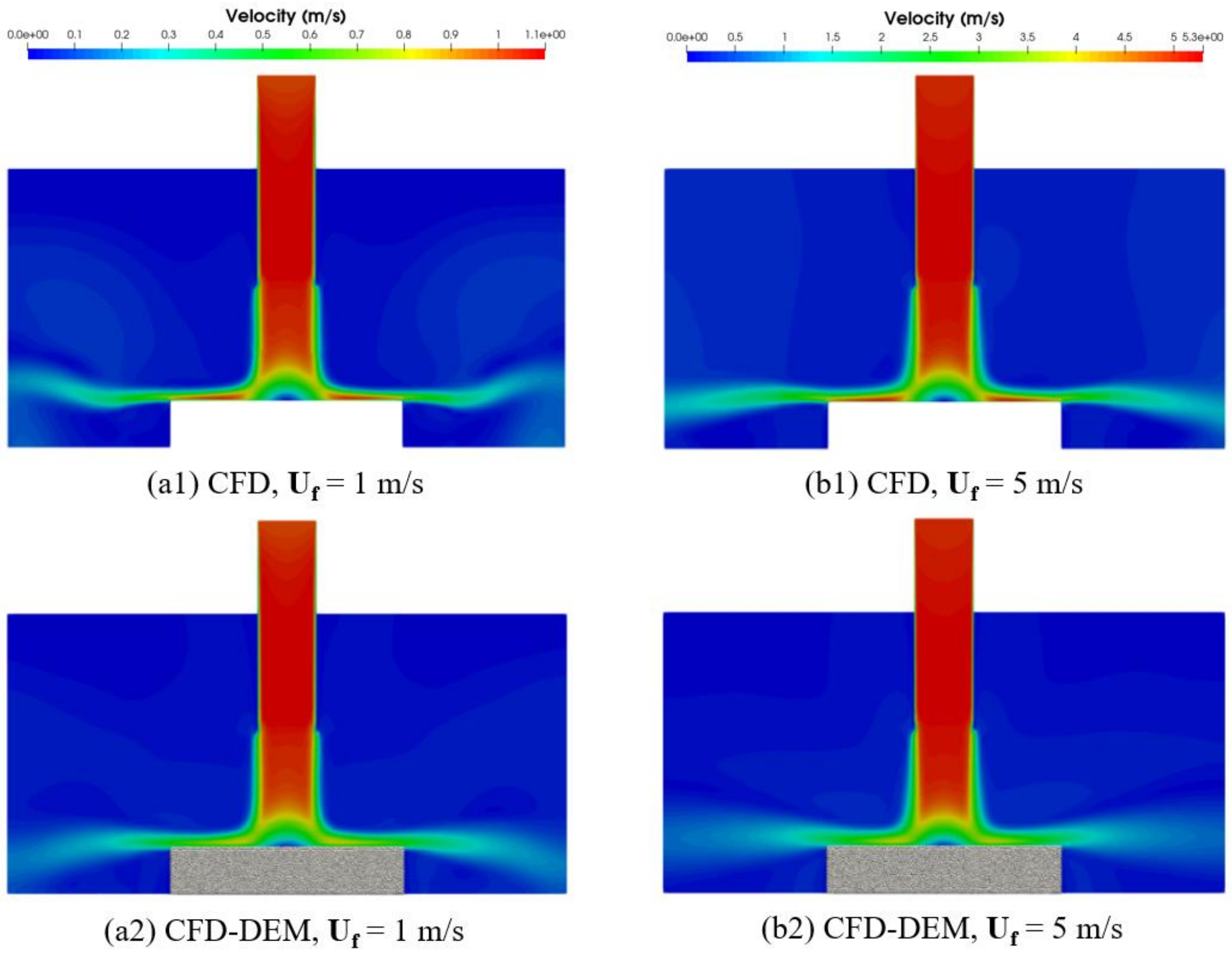

Fig. 15 Velocity contour from CFD and unresolved coupled CFD-DEM simulation of erosion test rig.

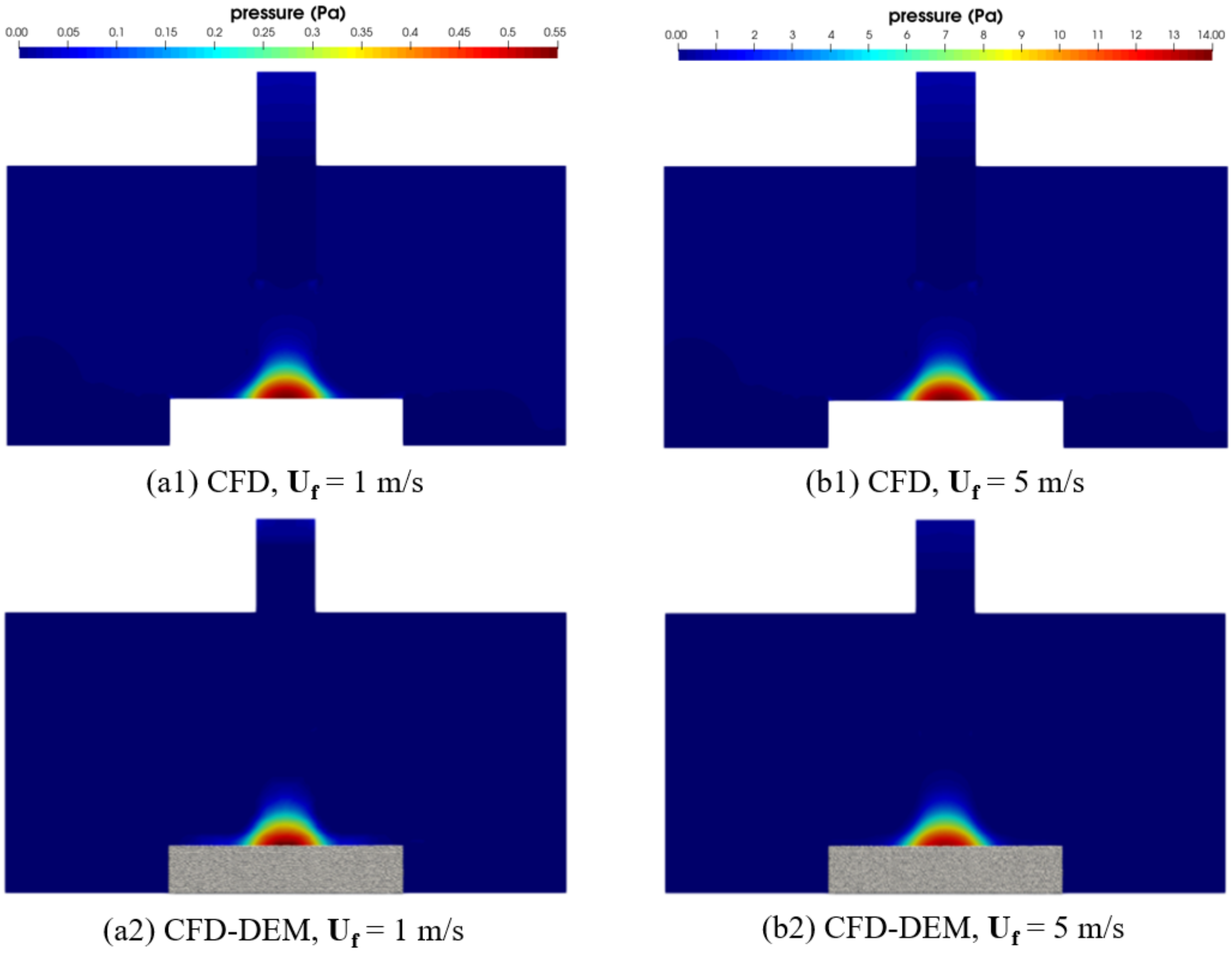

Fig. 16 Pressure contour from CFD and unresolved coupled CFD-DEM simulation of erosion test rig.

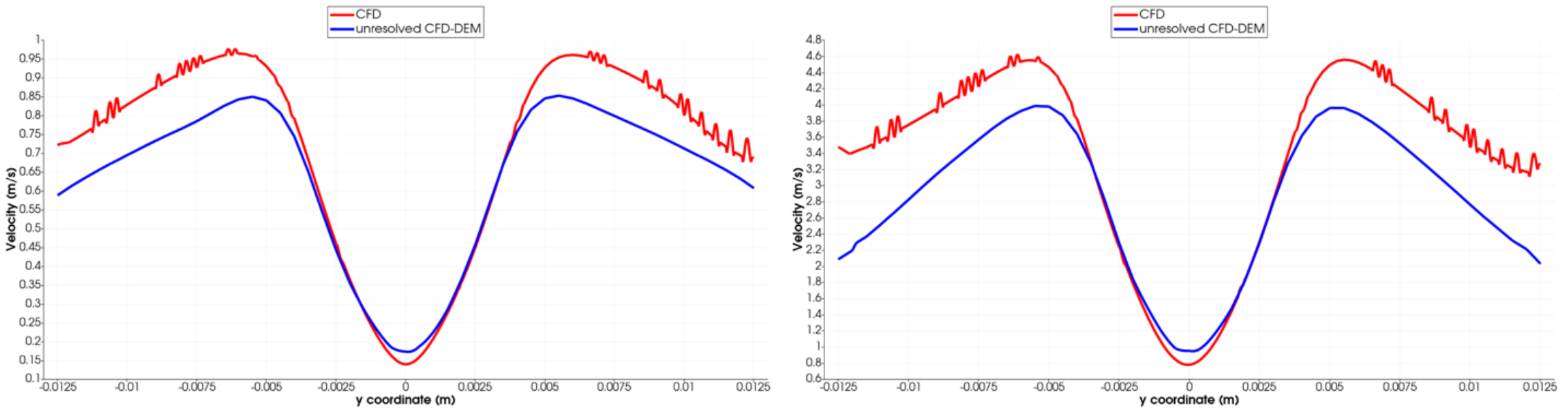

(a) $\mathbf{U_f}$ = 1 m/s, t =0.5s (b) $\mathbf{U_f}$ = 5 m/s, t =0.3s

Fig. 17 Velocity profile of fluid flow along a line located 0.5 mm above the sample surface in two cases of different discharge fluid velocity $\mathbf{U_f}$ =1 m/s and 5 m/s.

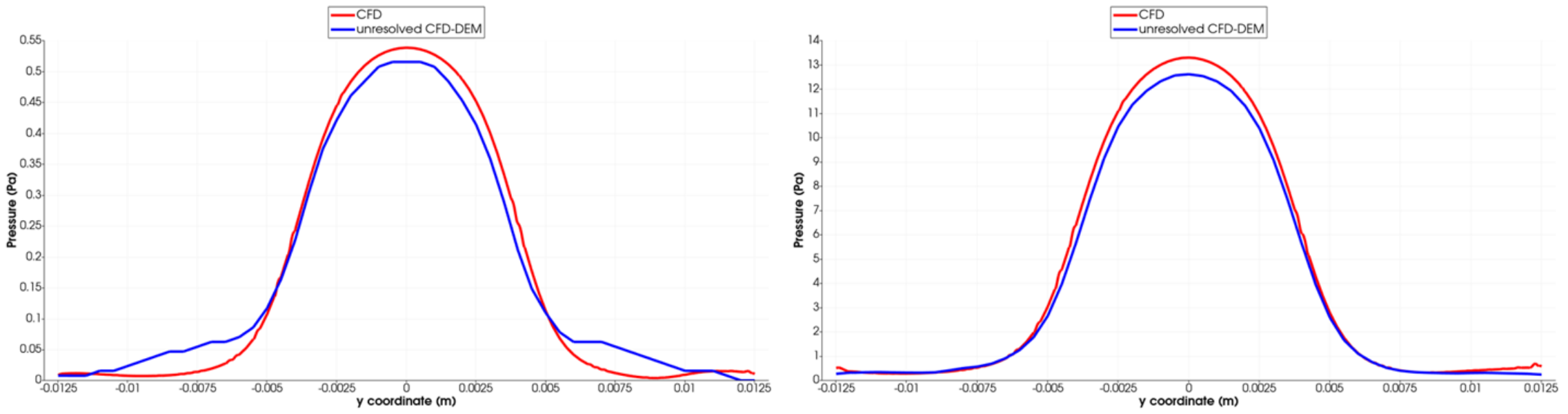

(a) $\mathbf{U_f}$ = 1 m/s, t =0.5s (b) $\mathbf{U_f}$ = 5 m/s, t =0.3s

Fig. 18 Pressure profile of fluid flow along a line located 0.5 mm above the sample surface in two cases of different discharge fluid velocity $\mathbf{U_f}$ =1 m/s and 5 m/s.

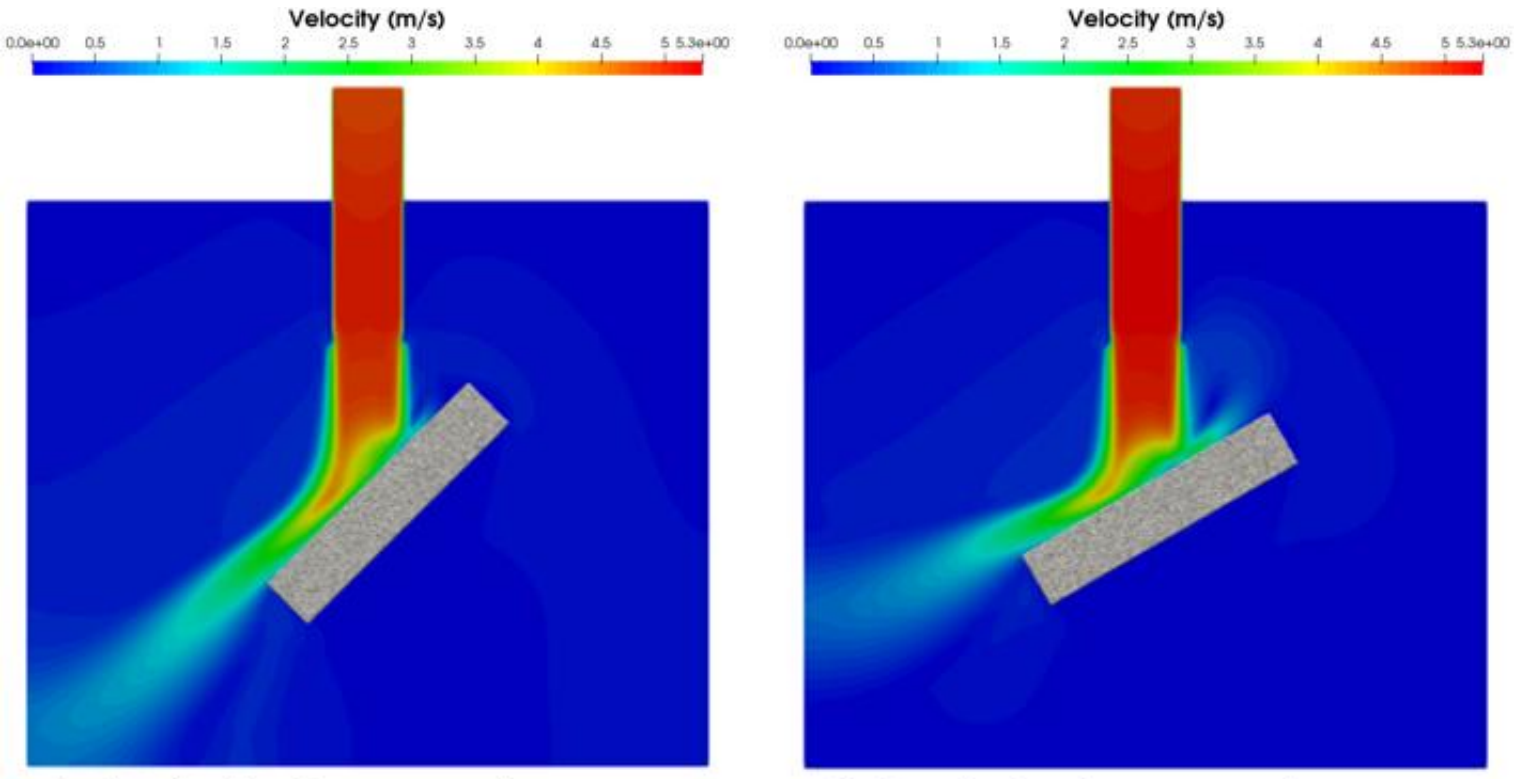

(a1) Velocity, impact angle = 45° (b1) Velocity, impact angle = 60°

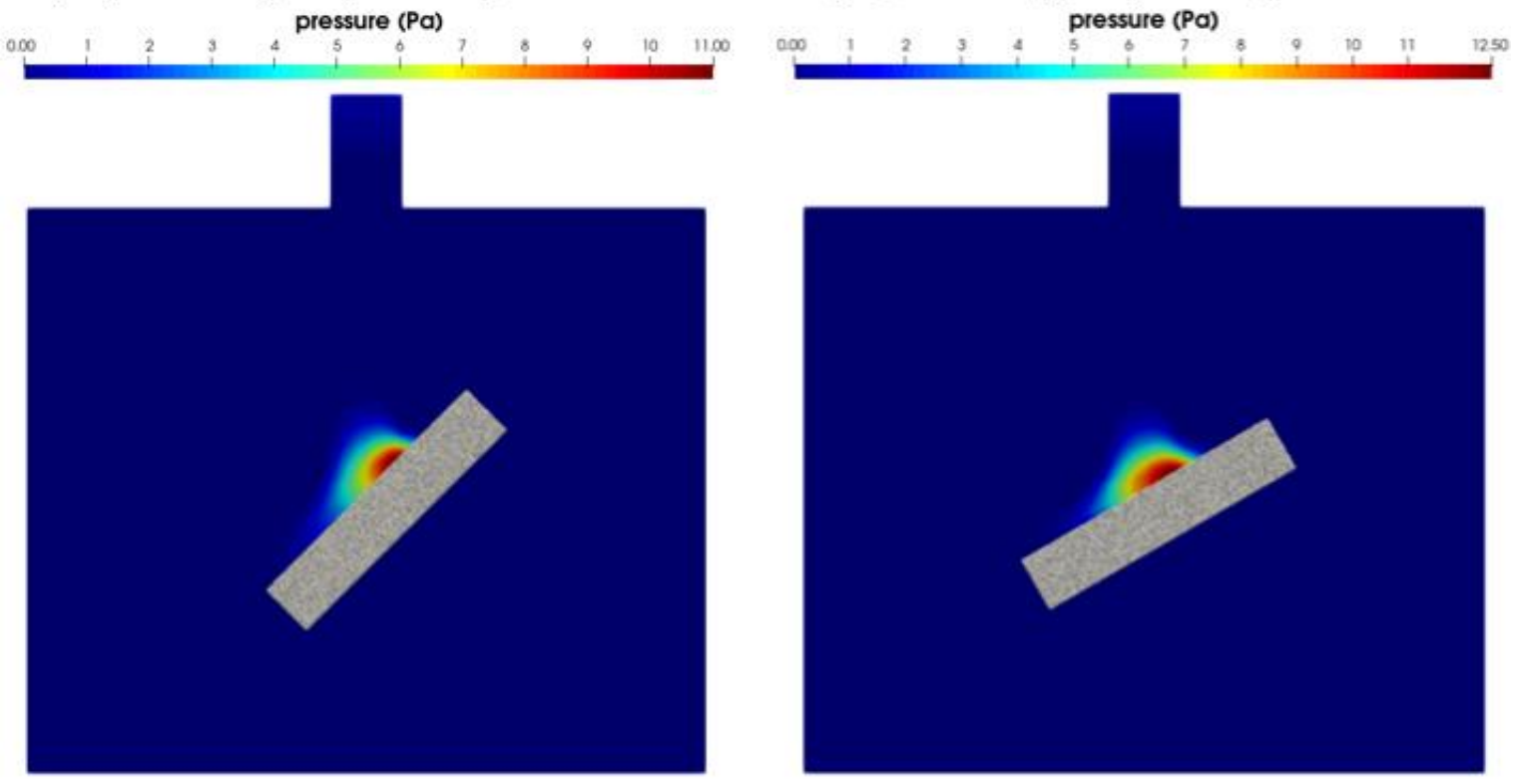

(a2) Pressure, impact angle = 45° (b2) Pressure, impact angle = 60°

Fig. 19 The velocity and pressure contour of erosion test rig configuration with oblique impact angle.

## 5.4. Erosive wear investigation

Building upon the previous section, an exploration of both (1) traditional approach using rigid continuous wall with Finnie erosion model [42, 43] and (2) the proposed full-discrete (or full particle-scale) model, will be presented in order to analyze erosion phenomena. An additional stream of “free-particles” (or sand particles) is inserted into the nozzle with initial velocity $\mathbf{U_p}$ and mixed with the water flow, which has a velocity $\mathbf{U_f}$. This particle-fluid stream subsequently collides with the “wall-particles”, as illustrated in Fig. 20.

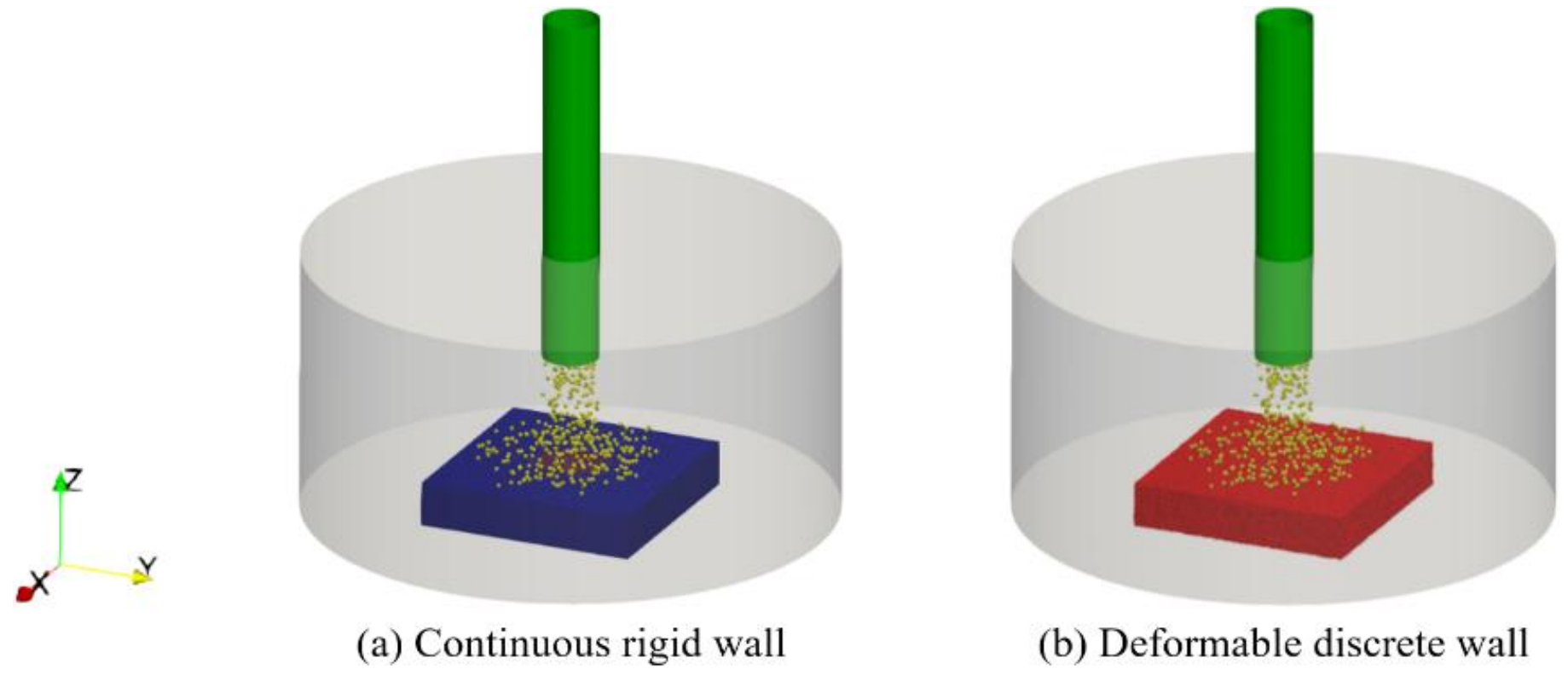

Fig. 20 Erosion simulation configuration of (a) traditional approach with continuous rigid wall, (b) the proposed model with deformable discrete wall.

The erosion process can be divided into two stages: pre-impact and impact. During the pre-impact stage, the carrier fluid transports particles toward the solid wall, while during the impact stage, particles collide with the solid wall, leading to erosive wear. An exclusive study on the impact stage provides a quick prediction of the erosive wear rate and pattern, as well as preliminary estimations of bond properties, time step, and particle size. Fluid flow provides momentum and directs erodent particles toward the wall during the pre-impact stage, and also carries them away from the eroded solid surface. Two approaches, referred to as the “dry test” and the “wet test”, are used. A “dry test” is defined as a simulation case that neglects the effect of fluid and concentrates solely on the interaction between particles and the solid surface wall during the impact stage. In contrast, a “wet test” takes into account the influence of fluid, which is typically present and important in practical scenarios. A real-life phenomenon corresponding to the “dry test” is sand blasting, while the “wet test” is erosion in the oil pipelines. The “dry test” simulation requires a particle solver (LIGGGHTS) and the implementation of Finnie or Oka erosion model. Meanwhile, the “wet test” involves a fluid solver (OpenFOAM), a particle solver (LIGGGHTS) and a coupling framework (CFDEM). The study in references [35, 45, 46] have successfully validated the Oka erosion equation through complex experimental systems. However, a drawback of the Oka model is its inclusion of numerous parameters which takes time for the calibration process. This is a challenge in practical applications, particularly when studying different wall materials, sand particles compositions, and flow conditions. On the other hand, Finnie model, one of the earliest and most fundamental models for recent developments, performs exceptionally well for erosion prediction for ductile materials with very few parameters. Furthermore, the simplicity of the Finnie model makes it suitable for quantitative validation in the initial testing of the proposed full particle model. In the full discrete model, after a period of operation time corresponding to a number of collisions, the bonds between wall-particles are broken. The propagation of broken bonds represents crack propagation as predicted in Fracture Mechanics, which subsequently leads to the removal of wall-particles. It is the erosive wear caused by the impact of sand particles on a (discrete) wall. The influence of fluid in dry and wet test is highlighted in the comparison between the “dry test” and “wet test”. All sand-particles in the current study are of the same size, while practical applications may contain particles with a wide range of size. A summary of sub-models and complexity of simulation test cases is shown in Table 2. The complexity and the demand of computational resources increase with the multiphase and multi-scale modelling of coupled CFD-DEM and inclusion of DEM bond network.

It has been observed from the erosion tests in our laboratory [47] as well as other references [35, 45, 46] that erosion often takes hours to occur. Because this is the first fundamental study, the computational time step in all of simulation will be chosen small enough to ensure the stability of numerical solution, as well as to accurately model the interaction between particles at microscale. Most of the time steps are of the order of 0.1 micro-second (1e-7 s). Due to computational resource limitations, but to maintain the generality of the test case, the particle mass flow rate is significantly increased, allowing erosion to be observed after 500,000 to one million computational time steps. In addition, the bond strength of “wall-particles” is reduced so that the bonds can be broken easier and shorter within that amount of time step. The fluid flow velocity used in this test is smaller than the reference (1-10 m/s compared with 10-30 m/s). Consequently, the bond properties do not represent any real materials, instead, they are used solely for the demonstration of the proposed model. The bond strength of real materials is often hundreds to thousands of times greater than in the current example, so the time required for erosive wear to occur would be much longer than demonstrated here. Moreover, the Finnie [42, 43] and Oka [48, 49] erosion equation use different parameters, so it is difficult to compare the results quantitatively. Instead, the qualitative evaluation focusing on important physics is employed throughout this section. It is also noted that the observable wear depth is highly dependent on the wall-particles size, as the wall-particles size represents the smallest unit of material removal. The smaller the wall-particle size is, the higher the number of wall-particles increases. In turn, it will require higher computational resources. In this study, the wall-particles are a few hundred microns in size, so only wear depths larger than this size are observable.

Table 2 Erosion simulation test cases.

| **Simulation Model** | **Continuous rigid wall (Finnie)** | | **Deformable discrete wall (current work)** | |
|---|---|---|---|---|
| | **Dry test** | **Wet test** | **Dry test** | **Wet test** |
| Fluid | No | Yes | No | Yes |
| Pre-impact stage | No | Yes | No | Yes |
| Impact stage | Yes | Yes | Yes | Yes |
| Coupling | No | Two-way Fluid-particles | No | Three-way Fluid-particles-structure |
| Erosion prediction | Finnie erosion equation | Finnie erosion equation | Count number of removed wall particles | Count number of removed wall particles |
| Eroded surface update | No | No | Yes | Yes |
| Complexity | Increasing from left to right | | | |
| Detail of erosion process included in the simulation | Increasing from left to right | | | |

In summary, after conducting numerous test cases, the material properties, bond strength, contact model parameters and fluid flow velocity are adjusted to be compatible with the computational resources available in our laboratory and to ensure numerical stability and convergence. The common parameters used throughout this study are listed in Table 9, Appendix A. Although the material and impact conditions are reduced in this study compared with the references [35, 45, 46], important features such as surface morphology (W- and U-shape) are preserved. The observed results reveal an eroded pattern on the wall surface, as seen on the z-plane at 5 mm, and the depth of wear is shown on the central sample cross section y-plane.

❖ **Finnie erosion model**

The only results observed from the use of erosion equation is the erosive wear pattern, which is proportional to the wear depth. To visualize the wear depth, we need to compute the volume loss based on the erosive wear rate and then update the mesh. However, the erosion model relies on semi-empirical coefficients, so the estimated volume loss requires an accurate calibration process. Instead, the qualitative comparison based on pattern is sufficient at this stage, while the wear depth will be shown by the full discrete model.

Two types of eroded surface profile are found from the combined experimental and numerical investigation: “W-shape” is caused by particle size smaller than and or equal to 400 µm, whereas “U-shape” is caused by particle size larger than 400 µm [46]. However, based on the results presented in this study, it is argued that the critical particle size, which differentiates between large and small particles, is relative depending on many factors, such as target surface materials, particle properties, fluid flow and impact conditions.

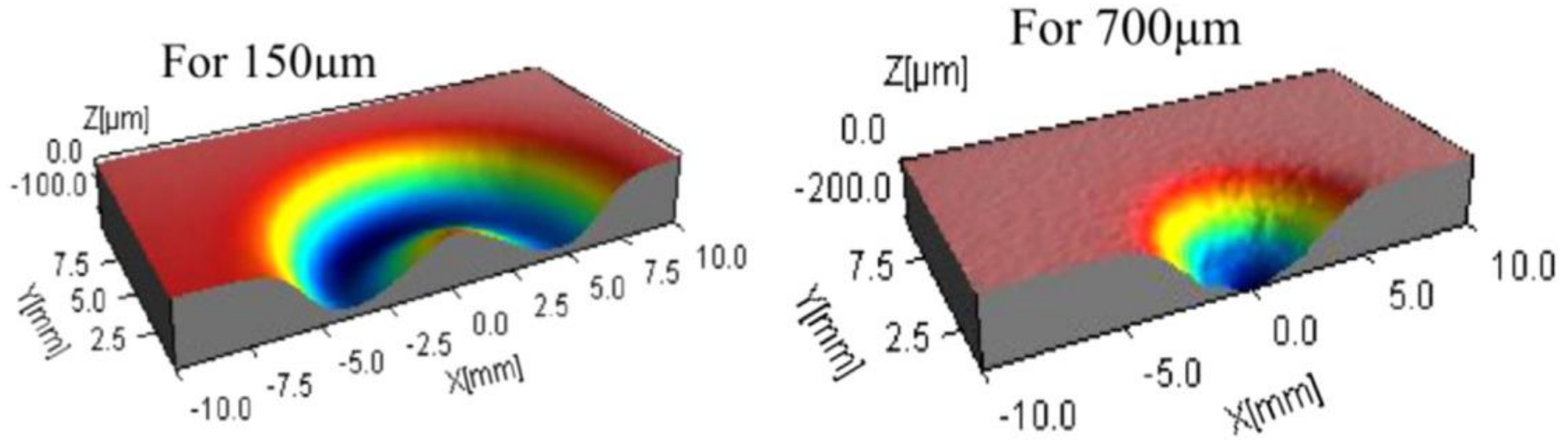

Fig. 21 Deformation of eroded surface caused by particles with different sizes: left (“W-shape”) with sand particle size smaller than 400 µm and right (“U-shape”) for others above 400 µm [46].

From the graphical representation in Fig. 21, it is evident that W-shape of eroded surface with small particle is shallower, but wider than the U-shape. The reason explained by the authors [46] is that the small particles are easily affected by fluid flow to move away from the center point of the nozzle, resulting in the spread of the wear area across the surface. Meanwhile, larger particles, with their greater momentum, continue moving toward the surface, causing a deeper wear.

In the dry test without fluid in this study, we firstly reproduce the erosive wear caused by particle stream on the eroded surface by Finnie model for two particle types of small and large, each type with two sizes:

- small particle with radius $r = 150$ um; and $r = 200$ um
- large particle with radius $r = 350$ um; and $r = 500$ um

As the first observation for qualitative evaluation shown in Fig. 22 and Fig. 23, the Finnie erosion model results in similar wear pattern with Oka: small particles (radius 150 um, 200 um) cause the W-shape erosive wear profile, while large particles (radius 350 um, 500 um) produces the U-shape. Upon close examination of the case of small particles, in both sub-cases of 150 μm and 200 μm, a small circular zone around the center of the test sample exhibits minimal wear, indicating that it is nearly touched by particles. From the center to a radius distance approximately 25% of the half-sample size (~ 3 mm), the circular wear pattern gradually increases, reaching its maximum value at 3 mm. Beyond this point, the wear starts to decrease. As the size of the particles increases, a noticeable effect is observed in the erosion pattern on the surface of the test sample. Specifically, the ring shape formed around the center of the sample begins to move towards the center as the particle size is increased. Since larger particles possess higher kinetic energy, they remove material more quickly and leave a deeper scar when colliding with the surface of the sample. As erosion progresses, a concave shape forms on the surface around the center point, which, in turn, attracts more particles due to gravitational and frictional forces. This process continues until the entire surface around the center point has been eroded away, resulting in a fully eroded surface with a central depression as in the case of large particle ("U-shape"). Based on the data visualized in these figures, it is apparent that there is a transition from a "W-shape" eroded profile, as illustrated in Fig. 22, to a "U-shape", as shown in Fig. 23, when particle size increases. The erosive wear profile is tracked over time steps to observe the progression of erosion, as depicted in Fig. 24.

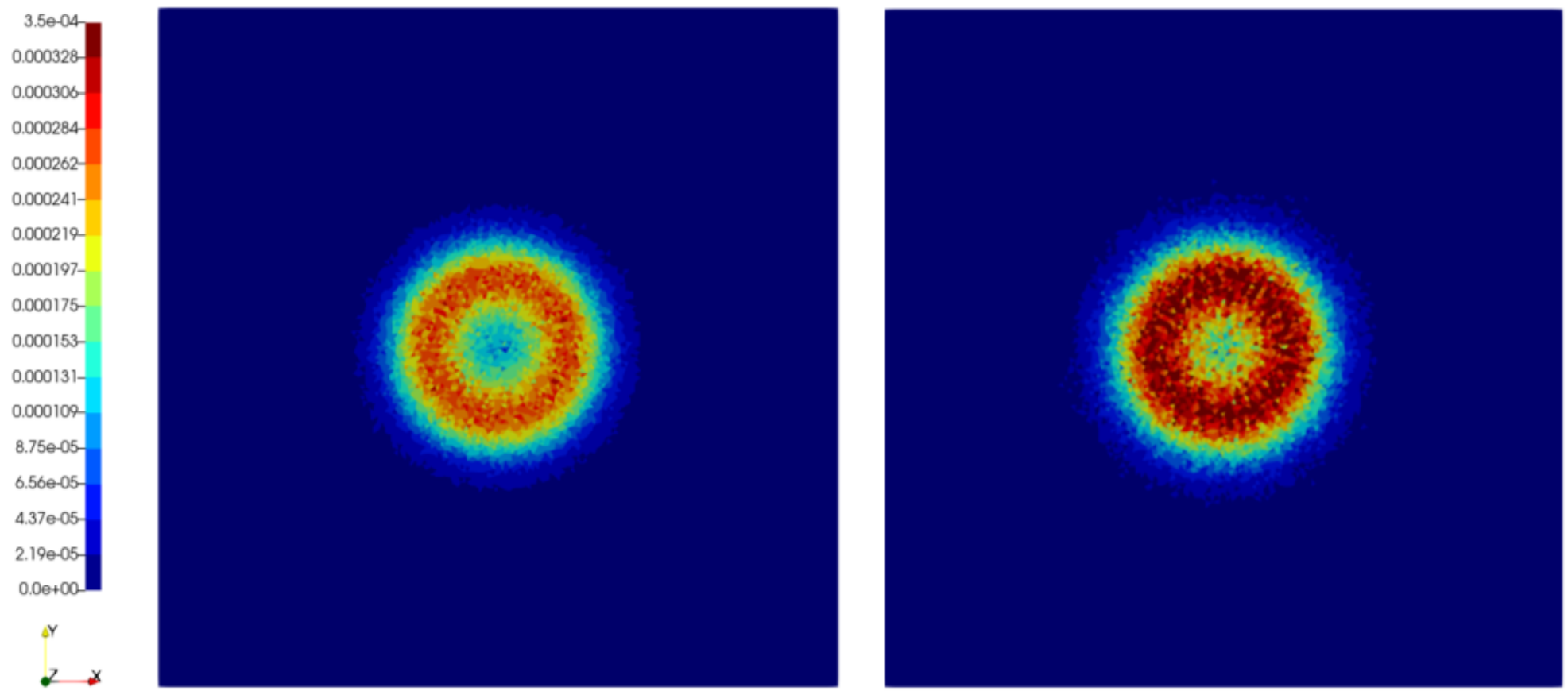

(a) DEM, Finnie, r = 150 μm (b) DEM, Finnie, r = 200 μm

Fig. 22 Erosive wear rate and pattern predicted by Finnie erosion model in case of small particles with size smaller than 300 μm after a million time steps and particle velocity $\mathbf{U}_f = 1$ m/s.

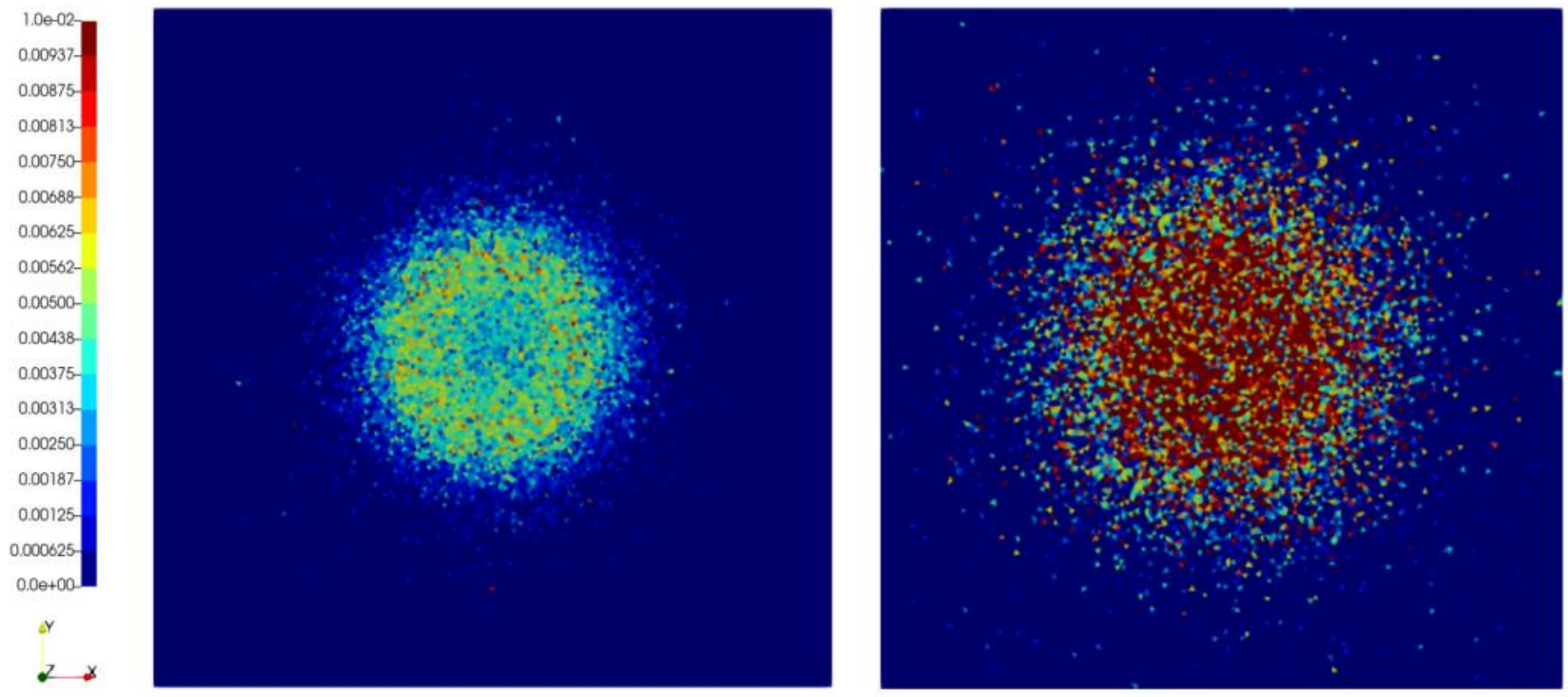

(a) DEM, Finnie, r = 350 μm (b) DEM, Finnie, r = 500 μm

Fig. 23 Erosive wear rate and pattern predicted by Finnie erosion model in case of small particles with size greater than 300 μm after a million time steps and particle velocity $\mathbf{U}_f = 1$ m/s.

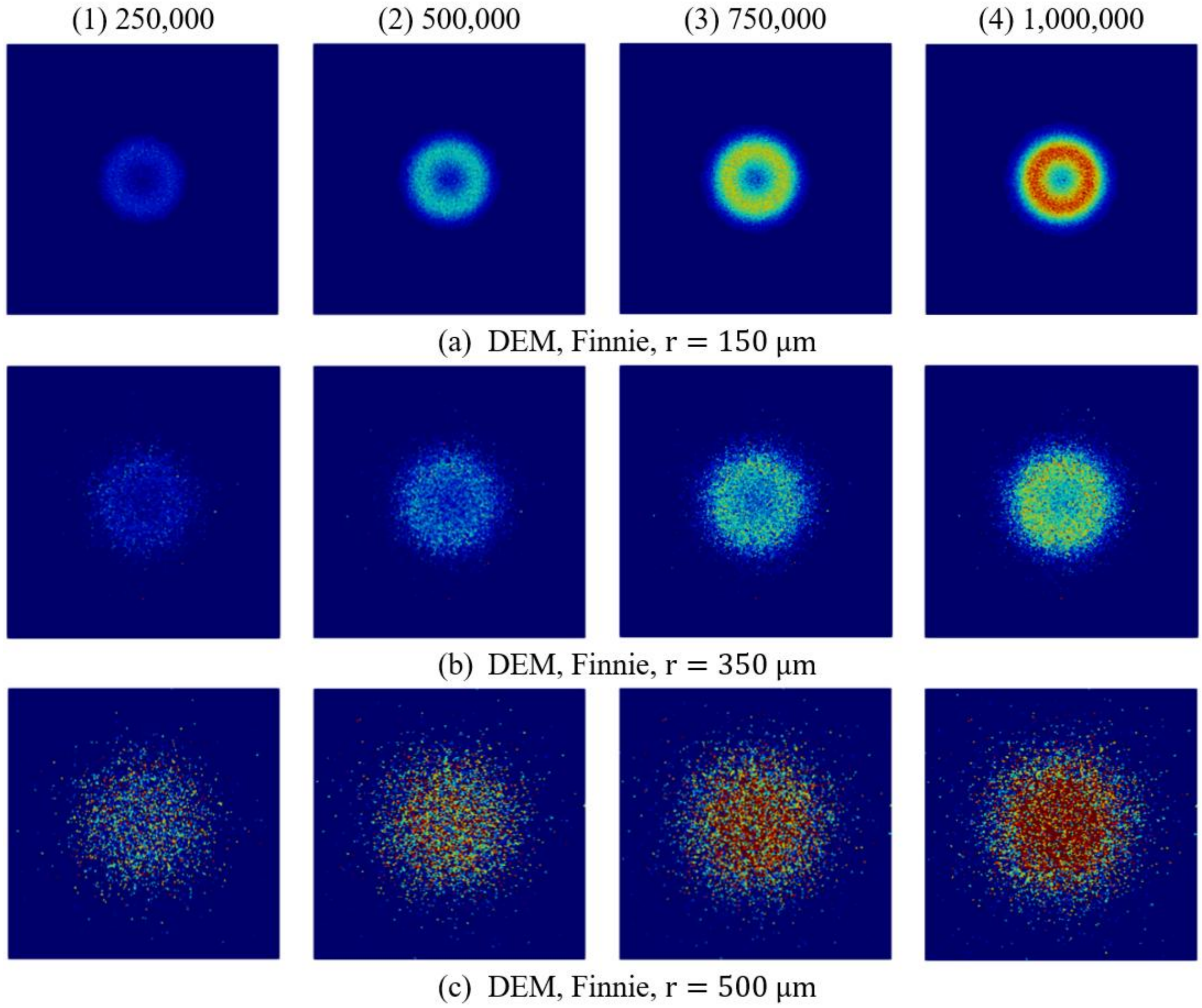

Fig. 24 Erosive wear profile caused by particle with radius 150 μm (row a), 350 μm (row b) and 500 μm (row c) over a number of time steps (1-4) and particle velocity $\mathbf{U_f}$ = 1 m/s.

After qualitatively validating the erosive wear profile with Oka erosion equation as well as experimental observation from the mentioned references, the effect of fluid is investigated here. Fig. 25 shows the comparison between purely discrete element simulation and coupled CFD-DEM. A visual analysis of Fig. 25 reveals that the fluid stream reduces the maximum value of erosive wear while keeping the similar pattern on eroded surface. Although the maximum value of wear is reduced in the center of plate, which is the impacting location of nozzle stream, the wear is apparently spread out in the whole surface, even to the further place near edges of the plate. The reason is that the fluid flow continues to carry away the particles after impacting the plate; thus, even at the further corner of the plate, the particles still collide with the sample surface during their movement.

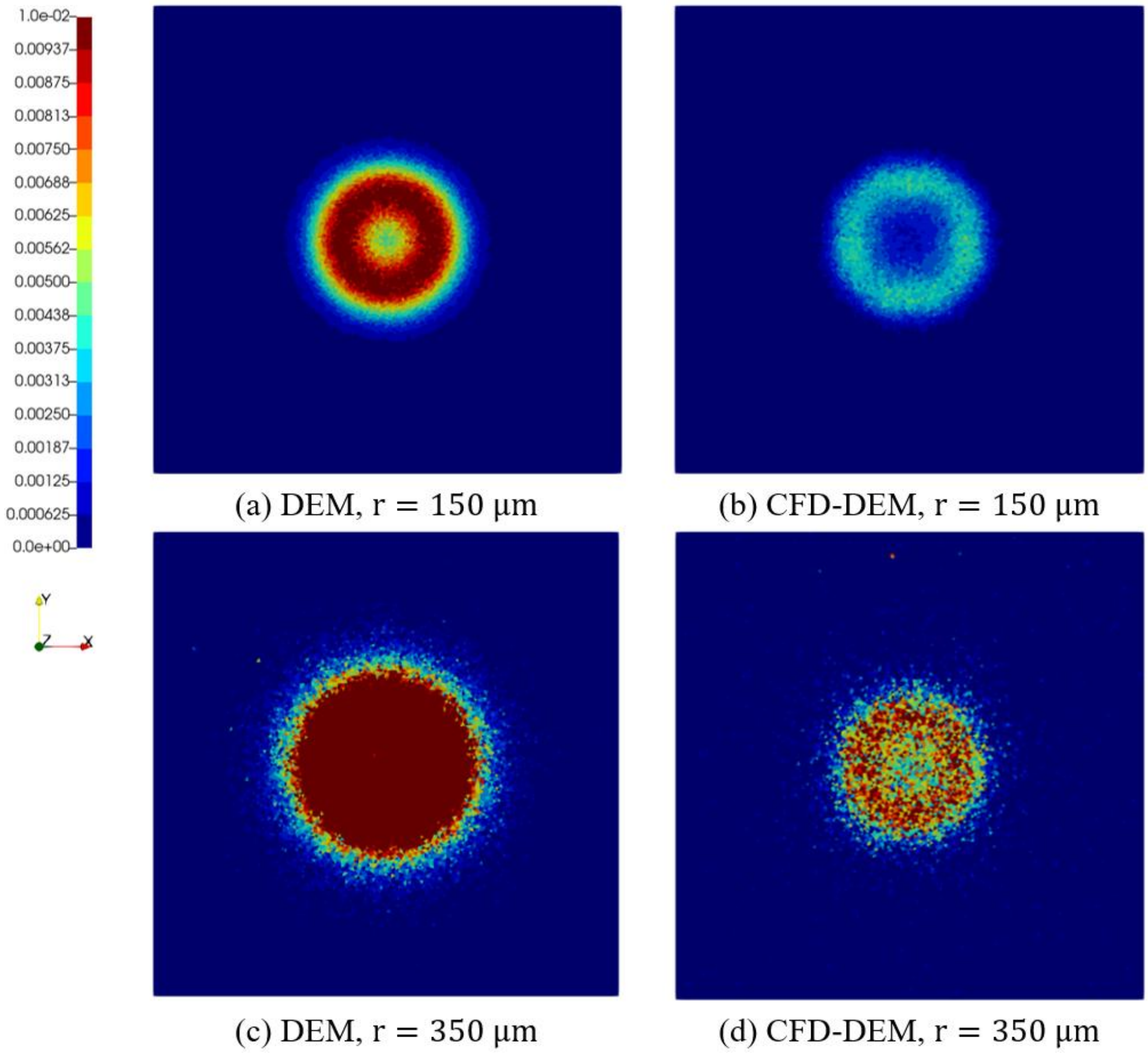

Fig. 25 Surface eroded profile predicted by Finnie erosion equation. Fluid velocity $\mathbf{U_f}$ = 5 m/s and particles stream velocity $\mathbf{U_p}$ = 1 m/s. The dry test with pure DEM and wet test with coupled CFD-DEM simulation in two cases of small particle r = 150 µm and large particle r = 350 µm after 500,000 time steps.

### ❖ Full discrete (or full particle-scale) model

The full particle-scale model has the following sub-models:

- DEM using bond to model materials.
- DEM tracks particles.
- Coupled CFD-DEM to model the mixture flow of fluid-particle.

It requires an additional contact model for interactions between free sand particles and wall particles, as well as between wall particles. For simplicity, the same contact model parameters listed in Table 9 are used without losing the generality of the test case. The full particle-scale model with a deformable discrete wall reveals no eroded pattern as in the case of the Finnie erosion equation [42, 43]. Instead, the eroded pattern comes from the deformation of discrete wall and removal of wall-particles. When a number of particles are removed from the wall surface, the remaining particles inside the thickness of the test sample (plate) represent the depth of erosive wear. In other words, the erosive wear depth is calculated as the distance from the original surface to the deepest remaining particles in the thickness direction (z-axis). The greater this distance, the more severe the erosive wear. One third from the bottom of test sample is fixed, therefore, the scale of the wear depth is only from 0 to 3.5 mm, where value 0 represents zero wear and value 3.5 mm represents the maximum wear. Erosive wear is considered to be proportional to the number of removed discrete wall-particles.

Similar to the dry test in the case using Finnie model and continuous rigid wall, this section evaluates the performance of full particle scale model. The erosion pattern and wear depth in case of particle radius 150 µm and 350 µm are depicted in Fig. 26 - Fig. 29. Upon inspecting those figures, it can be concluded that the full particle scale model results in an eroded profile “W-shape” and “U-shape” for small and large particles, respectively. However, there is a little difference in the expected results compared with the traditional approach. The simulation using Finnie erosion model with a continuous rigid wall captures all erosive wear rates, even if they are very small, such as 0.01 or 3.5e-4. In contrast, the proposed full particle-scale model accounts for erosion only from the moment the first particle is removed. It implies that erosive wear is zero during the initial time steps although the discrete wall may experience deformation. It is clearly demonstrated that in the case of a 150 µm sand particle when it took a very long time, up to a million time steps, to observe deformation on the cross-section of the test sample. The Finnie case does not update the geometry, while the discrete wall accumulates deformation from wear throughout simulation process. The alteration of geometry affects the flow field and impact conditions. As a result, this contributes to the discrepancy between these two cases.

The final simulation is the most complicated one, which includes the coupled CFD-DEM for simulating the mixture flow of particles-fluid, discrete element with bond model to reproduce mechanical behavior of test sample. In this test, the effect of fluid flow is presented. The erosive wear pattern and depth in two case of particle 150 µm and 350 µm, representing small and large particle respectively, are shown in Fig. 30 and Fig. 31. It is similar to the case of Finnie model, the water flow reduces the effect of erosive wear. The wear depth and erosive wear width depicted in Fig. 30 and Fig. 31 clearly show the evidence for that effect. The fluid flow, particularly with turbulence, tends to carry particles far away from the center of impact. It is noted that the particles bounded after impacting the wall are also directed by the incoming water stream. As a result, they move away from the center and spread out on the surface. It is known that the tangential impact hardly removes particles, unlike normal impact in the initial configuration, but still leaves a wear scar. Furthermore, the vortices in turbulent flow also carry particles far away from the surface wall. The visualized data in Fig. 31 shows the severe erosion in case of particle radius 350 µm. This results from the scenario in which bonds connect particles in the discrete wall have low failure strength. Therefore, another test of increasing bond strength by double its value is conducted here. The results of the new case are illustrated in Fig. 32 with the new erosive wear depth greatly reduces from ~ 3 mm to ~ 1.7 mm.

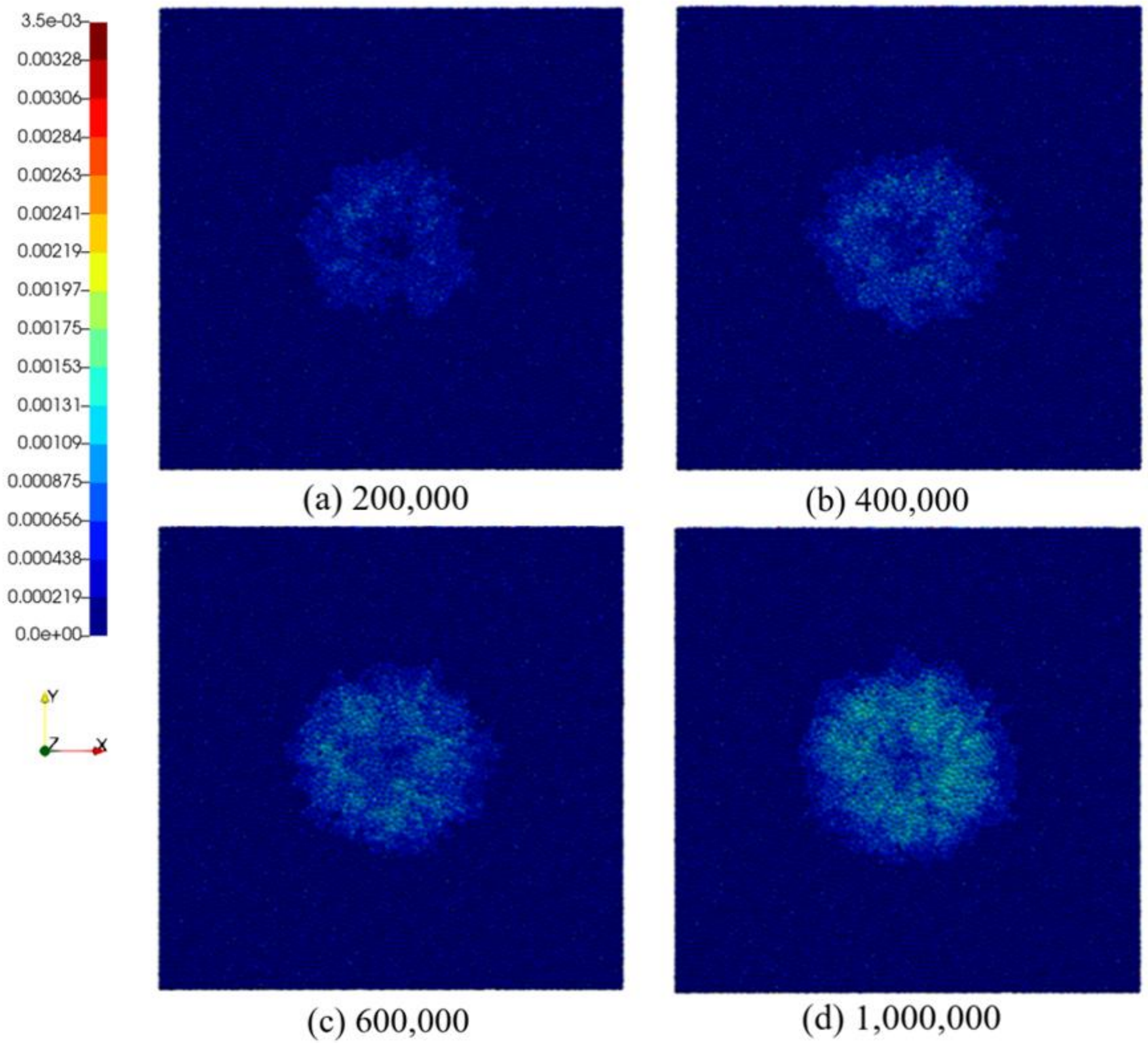

Fig. 26 Equivalent eroded profile caused by particles with radius 150 μm in the dry test without fluid. The wear depth is visualized by distance of wall particles to the surface in the thickness direction (z-axis) over a number of time steps (a-d). Particle stream velocity $\mathbf{U_p}$ = 5 m/s.

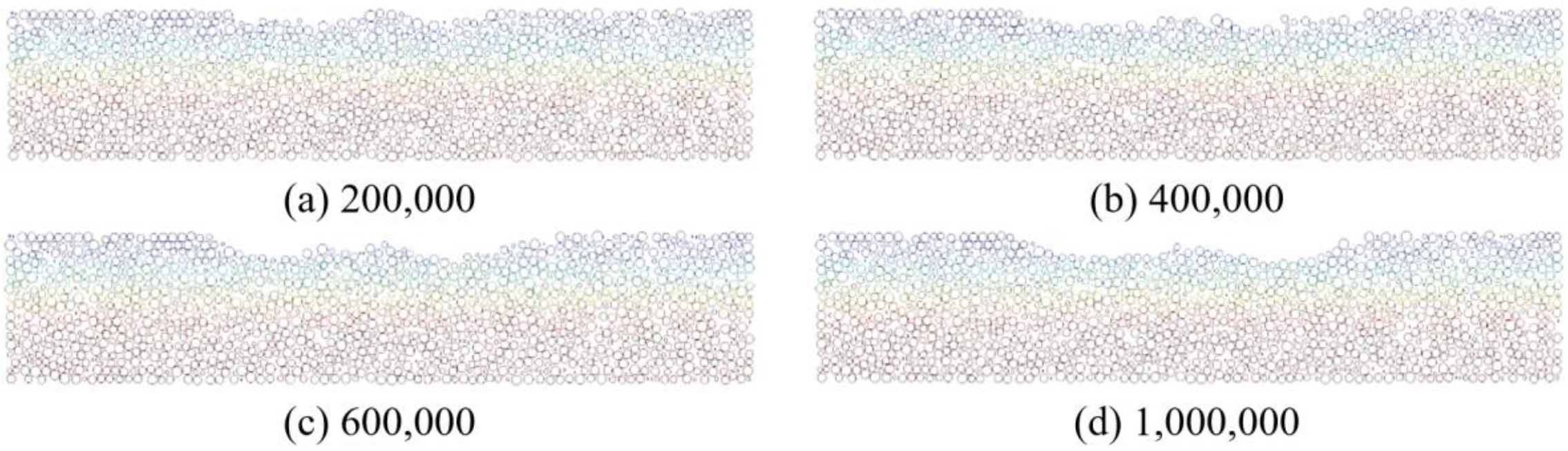

Fig. 27 Cross section of eroded plate caused by particles with radius 150 μm in the dry test without fluid. The wear depth is visualized by distance of wall particles to the top surface in the thickness direction (z-axis) over a number of time steps (a-d). Particle stream velocity $\mathbf{U_p}$ = 5 m/s.

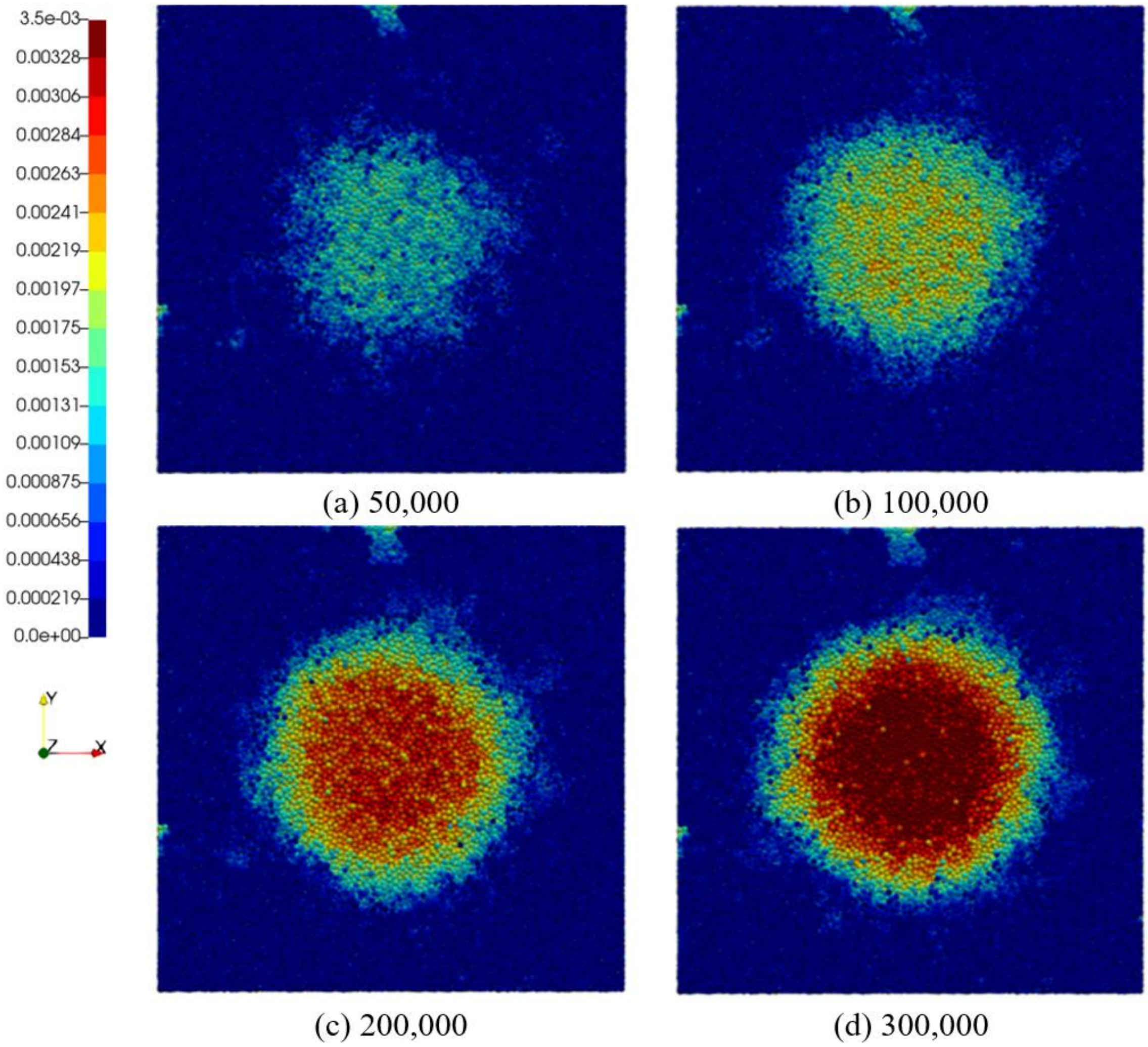

Fig. 28 Equivalent eroded profile caused by particles with radius 350 μm in the dry test without fluid. The wear depth is visualized by distance of wall particles to the top surface in the thickness direction (z-axis) over a number of time steps (a-d). Particle stream velocity $\mathbf{U_p}$ = 1 m/s.

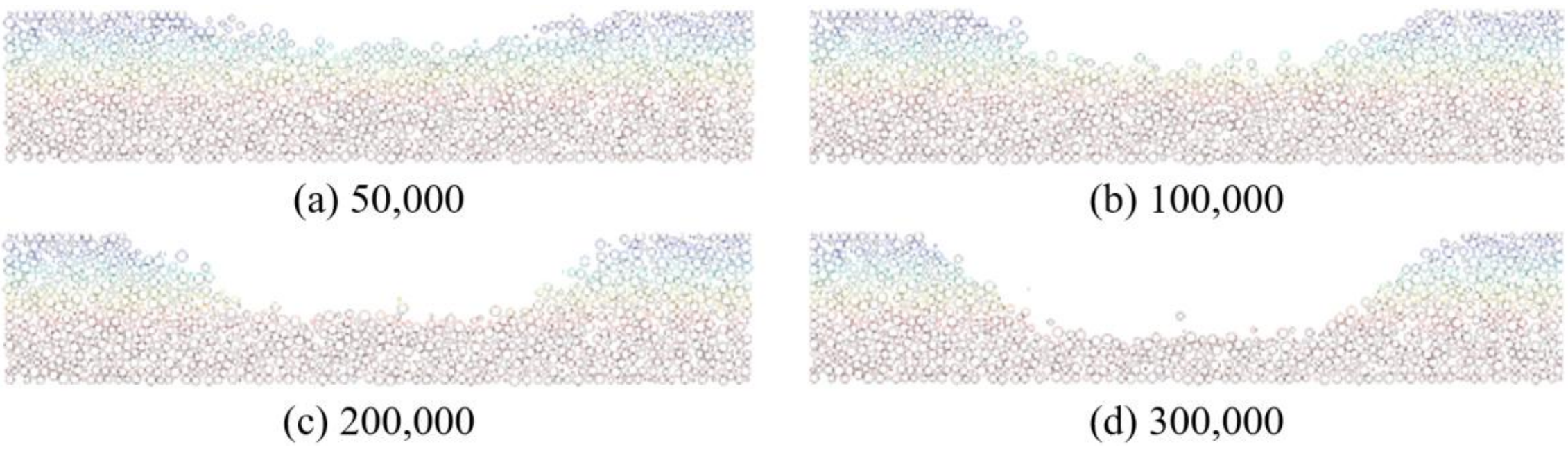

Fig. 29 Cross section of eroded plate caused by particles with radius 350 um in the dry test without fluid. The wear depth is visualized by distance of wall particles to the surface in the thickness direction (z-axis) over a number of time steps (a-d). Particle stream velocity $\mathbf{U_p}$ = 1 m/s.

As previously explained, the erosive wear is proportional to the removed discrete particles from wall. The influence of particle radius and particle velocity are summarized in Table 3 and Table 4. In terms of particle radius investigation (Table 3), it is noted that the number of removed particles, are increased significantly with the particle radius of 350 um. Meanwhile, the study of fluid flow velocity (Table 4) also shows a remarkable number of removed particle in the case of 10 m/s. Due to the excessively reduced bond strength, the materials of surface wall cannot withstand the impact of a high-speed particle-fluid stream. In extreme test cases, such as a particle radius greater than 350 μm and a high velocity of 10 m/s, the entire test sample could be completely damaged. Therefore, in those cases, the removed particles no longer represent the erosion, instead, this is the structural failure. This is also a nice feature of the proposed full particle scale model compared to the case using Finnie. The use of a rigid wall, which ignores surface deformation, is not able to capture the structural failure. Instead, it consistently results in an erosive wear rate value, which is sensitive and depends on a number of coefficients.

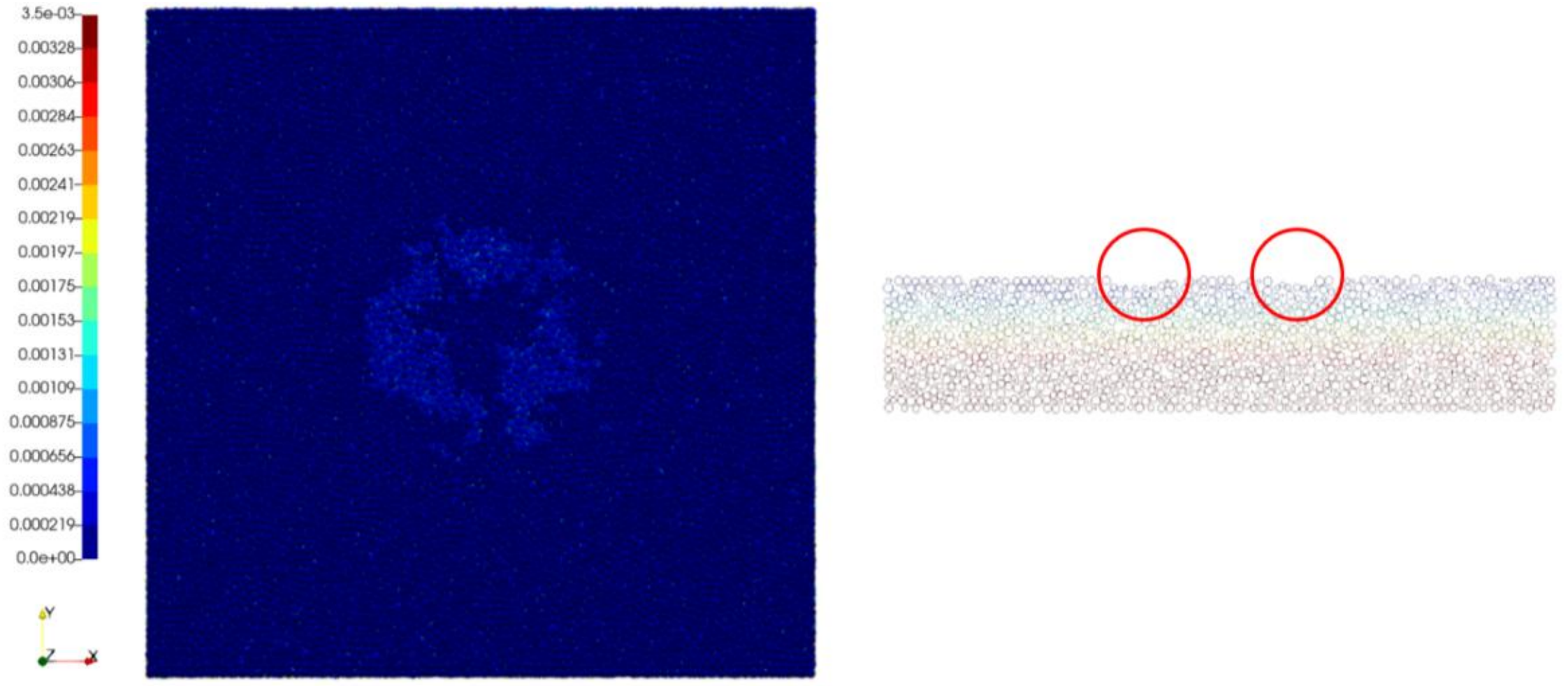

(a) Erosive wear pattern (b) Erosive wear depth viewed on cross-section

Fig. 30 Erosive wear pattern and depth predicted by full particle-scale model for small particles with radius 150 μm after a million time steps in the wet test with fluid.

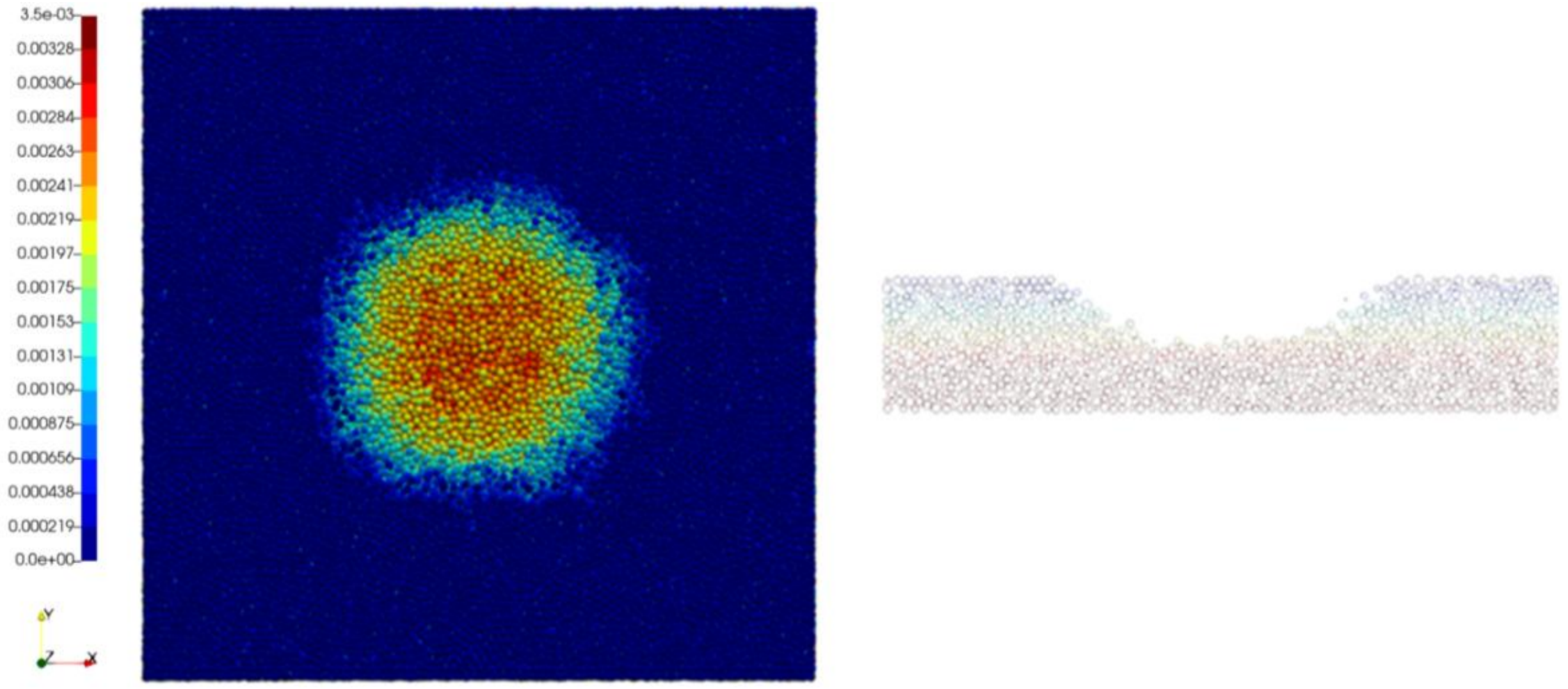

(a) Erosive wear pattern (b) Erosive wear depth viewed on cross-section

Fig. 31 Erosive wear pattern and depth predicted by full particle-scale model for large particles with radius 350 μm after 100,000 time steps in the wet test with fluid.

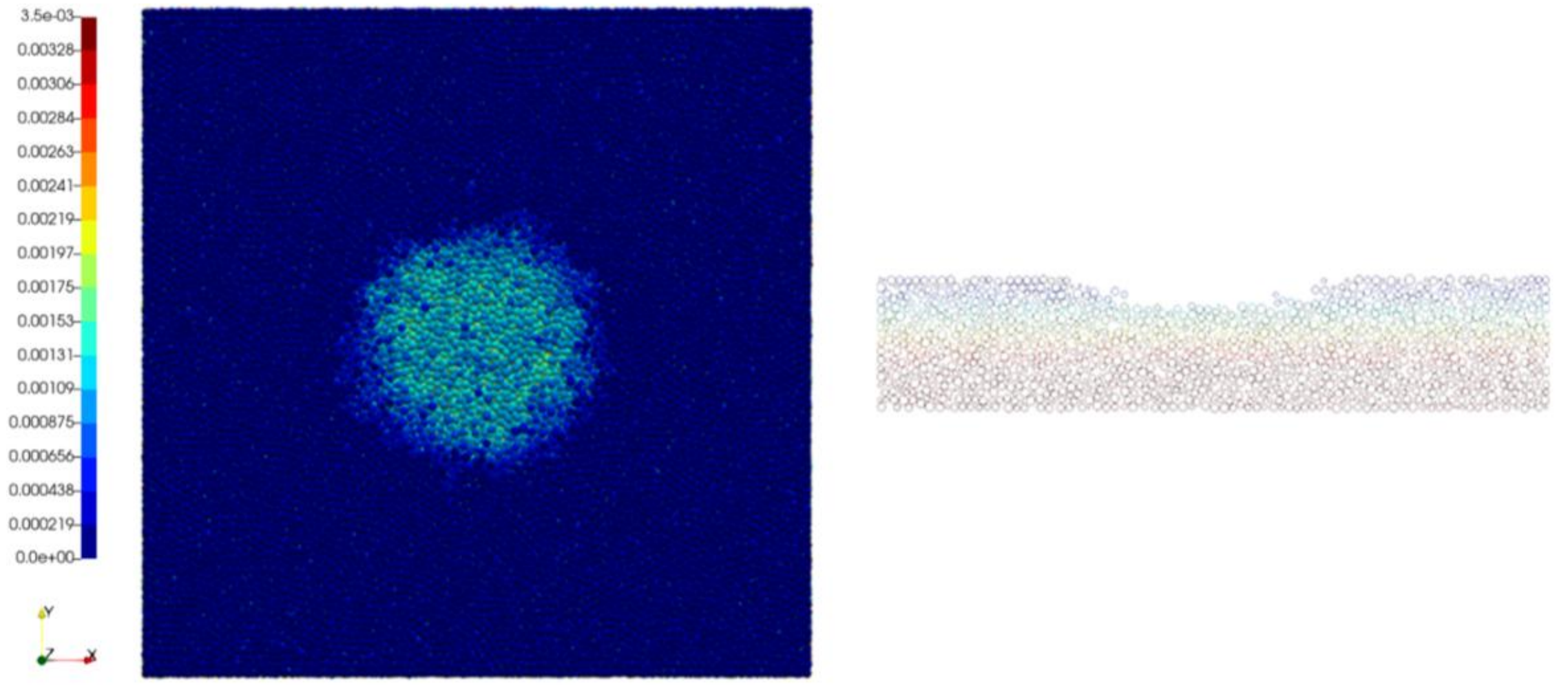

(a) Erosive wear pattern (b) Erosive wear depth viewed on cross-section

Fig. 32 Influence of bond strength on erosive wear pattern and depth caused by large particles with radius 350um after 100,000 time steps in the wet test with fluid, where the bond strength is doubled compared to the case in Fig. 31.

Table 3 Influence of particle size: Number of removed wall-particles after 200,000 time steps.

| **Particle radius (μm)** | 150 | 200 | 250 | 350 | 400 | 500 |
|---|---|---|---|---|---|---|
| **Number of removed wall-particles** | 11 | 2,139 | 5,425 | 13,589 | 14,248 | 17,477 |

Table 4 Influence of fluid flow velocity: Number of removed wall-particles in case of particle radius 150 um after 200,000 time steps.

| **Fluid flow velocity (m/s)** | 5 | 10 |
|---|---|---|
| **Number of removed wall-particles** | 11 | 10,388 |

In order to study the effect of number of bonded particles used for modelling the deformable discrete wall, three different cases are conducted with 100000 particles ($d_{avg}$~375 μm), 150000 particles ($d_{avg}$~300 μm) and 200000 particles ($d_{avg}$~270 μm). The bond failure strength is also increased to 5e5 Pa to allow erosion to occur without causing structural failure. The sand particle size is 200 μm. As previously explained, erosive wear begins when the first "wall-particles" is removed, and its value predicted by full discrete model is the number of wall particles. Consequently, Fig. 33 demonstrates that the greater the number of particles is, the sooner erosive wear can be observed. Case (c) with 200,000 wall-particles results in a clear erosive wear pattern, while wear in case (a) with 100,000 wall-particles is almost negligible, because the large particles now is a cluster of smaller particles and are more difficult to be removed.

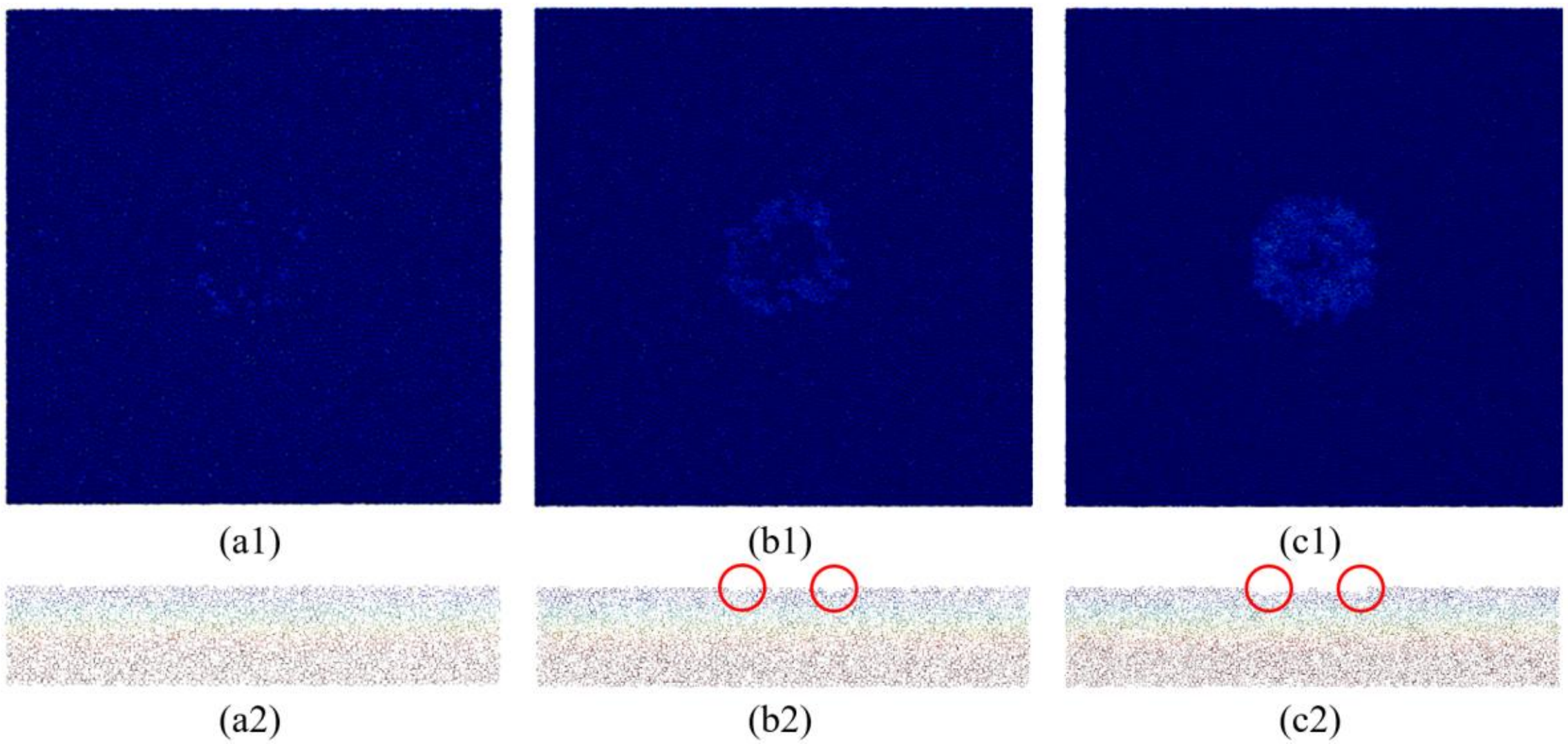

Fig. 33 Eroded profile and cross section observed after 600,000 time steps: (a) 100,000 wall-particles, (b) 150,000 wall-particles and (c) 200,000 wall-particles.

## 6. Conclusion

This paper aims to demonstrate that the replacement of a continuous solid object by a corresponding deformable discrete domain, generated by Random Close Packing (RCP) method, can reproduce the fluid flow response as in a CFD simulation. After that, a particle-based approach, which combines material modelling by DEM using bond model and multiphase flow simulation by coupled CFD-DEM, is proposed to simulate the flow over deformable object (FSI problem). It is referred to as “full discrete model” or “full particle-scale model” in this study, which emphasizes that the solid object is represented by micro-scale discrete elements. The benchmarks include laminar/turbulent flow, spherical/non-spherical (or blunt) object, resolved/unresolved coupled scheme, combination of rigid and discrete solid wall. Although the current coupled CFD-DEM scheme and DEM bond model cannot accurately reproduce the streamlines of flow near the solid wall as CFD simulation, it preserves the most important features of flow field in the whole computational domain like pressure and velocity contour, far field streamlines, near wall streamlines with little deviation as seen in benchmark examples. This approach is particularly useful when the microstructure of materials in particle-fluid flow plays a crucial role but difficult to be described by a single continuum mechanics law. Furthermore, by inserting free particle streams, it seamlessly simulates the complicated three-way interaction between fluid, particles and structure in the erosive wear study. The more bond types, interaction forces models and coupled CFD-DEM schemes are developed, the more types of materials and debris flow conditions it is expected to simulate. With further investigation and validation, the proposed approach may serve as an alternative numerical tool to simulate fluid structure interaction (FSI) or erosion phenomena, which also allows the study of complicated effect of materials microstructure in practical problem, such as materials with initial defects.

## Appendix A

Table 5 Fluid properties

| Parameters | Value |
|---|---|
| Density ρ ($\frac{kg}{m^3}$) | 1000 |
| Kinematic viscosity ν ($\frac{m^2}{s}$) | 1.0e-6 |

Table 6 Simulation configuration of flow over sphere

| Parameter | Value |
|---|---|
| Sphere diameter (D) | 0.02 m |
| Length (L1) | 0.2 m |
| Length (L2) | 0.4 m |
| Width (W1) | 0.2 m |
| Width (W2) | 0.2 m |
| Height (H1) | 0.2 m |
| Height (H2) | 0.2 m |
| Fluid velocity (U) | 0.005 m/s |
| Reynolds (Re) | 100 |
| Laminar / Turbulent | Laminar |

Table 7 Simulation configuration of flow over square cylinder

| Parameter | Value |
|---|---|
| Square cylinder (x, y, z) | 0.02 m x 0.02 m x 0.2 m |
| Length (L1) | 0.2 m |
| Length (L2) | 0.4 m |
| Width (W1) | 0.2 m |
| Width (W2) | 0.2 m |
| Height (H1) | 0.1 m |
| Height (H2) | 0.1 m |
| Fluid velocity (U) | 0.25 m/s |
| Reynolds (Re) | 5000 |
| Laminar / Turbulent | Turbulent |

Table 8 Simulation configuration of flow from nozzle impacting a plate.

| Parameter | Value |
|---|---|
| Plate (length x width x height), L x L x H | 25 x 25 x 5 (mm) |
| Nozzle diameter (D) | 6.4 (mm) |
| Distance from nozzle to plate (d) | 12.7 mm |
| Discharge fluid velocity (**U** or $\mathbf{U_f}$) | 1 m/s |
| Turbulent / Laminar | Turbulent |

Table 9 Bond properties used in erosion simulation.

| Parameter | Value |
|---|---|
| Fluid velocity (U, m/s) | 1 – 5 – 10 –15 |
| Turbulent / Laminar | Turbulent |
| Bond diameter (μm) | 100 |
| Bond Young's modulus (Pa) | 1.0e7 |
| Bond Shear modulus (Pa) | 3.8e6 |
| Critical normal stress (Pa) | 1.0e5 |
| Critical shear stress (Pa) | 1.0e5 |
| Contact Young's modulus (Pa) | 1.0e9 |
| Poisson ratio | 0.3 |
| Coefficient of restitution | 0.5 |
| Coefficient of friction | 0.1 |
| $k_{Finnie}$ (Finnie coefficient) | 1.0e-5 |
| Coupled CFD-DEM data exchange | Every 10 DEM time steps |

## Appendix B

It can be clearly observed that there is a large difference in physical computing time between resolved and unresolved scheme illustrated in Table 10. In all coupled CFD-DEM simulations, the general setting is that each CFD time loop has 10 DEM time steps, which means the information is exchanged between CFD solver and DEM solver happens every 10 DEM time steps. The time step in each simulation is adjusted to ensure the CFL number and numerical stability, hence there may be a little difference in time step value between sub-cases in a simulation case. As a result, the simulation time achieved with the same amount of computer resources is affected by the complexity of individual cases, including time step value, turbulent/laminar flow, number of particles, etc. In the flow over a stationary object like benchmarks in Section 5, we could save the resources by fixing all the particles without calculating bond reaction forces and updating particle's position. However, in the problem of Fluid Structure Interaction or erosive wear phenomena, the number of bonds in the discrete object has a great influence because we need to calculate the bond reaction forces, then update the movement of particles. In case that all bonds of a particle are broken, the particle is set to free, hence, the number of bonds is reduced, so is the computer resources spent for computing bonded contact. However, this particle can freely move and collide with the surrounding particles, so the resources spent for calculating unbonded contact becomes larger. It is difficult to predict the computer resources in those complex scenarios in advance.

Table 10 Computer resources spent in coupled CFD-DEM simulation to model the fluid flow over deformable discrete object.

| Case | Simulation | Number of particles | Number of CPU cores | Number of DEM time steps | Computing Time (hours) |
|---|---|---|---|---|---|
| Sphere (Laminar) | resolved | 1 | 96 | 150,000 | 7.9 |
| | resolved | 1,000 | 96 | 150,000 | 19.2 |
| | resolved | 5,000 | 96 | 150,000 | 26 |
| | resolved | 10,000 | 96 | 150,000 | 28 |
| | unresolved | 50,000 | 96 | 150,000 | 8.9 |
| Square Cylinder (Turbulent) | resolved | 5,000 | 192 | 150,000 | 48 |
| | resolved | 10,000 | 192 | 90,000 | 48 |
| | unresolved | 100,000 | 192 | 500,000 | 3.5-4.2 |
| Erosion test rig configuration without particle (Turbulent) | unresolved | 100,000 + | 48 | 1,250,000 | 47 |
| | unresolved | 100,000 + | 96 | 925,000-1,100,000 | 24-34 |

**Appendix C**

Mathematical models of unresolved coupled CFD-DEM schemes

- Set I (Model BFull):

$$\frac{\partial(\rho_f \varepsilon_f u)}{\partial t} + \nabla.(\rho_f \varepsilon_f u u) = -\nabla p - F_{pf}^{setI} + \nabla.\tau + \rho_f \varepsilon_f g \quad (4)$$

where:

$$F_{pf}^{setI} = \frac{1}{\Delta V}\sum_{i=1}^{n}\left(f_{d,i} + f_{\nabla p,i} + f_{\nabla.\tau,i} + f_i''\right) \quad (5)$$

$$f_{pf,i} = f_{d,i} + f_{\nabla p,i} + f_{\nabla.\tau,i} + f_i''$$

- Set II (Model A):

$$\frac{\partial(\rho_f\varepsilon_f u)}{\partial t} + \nabla.(\rho_f\varepsilon_f uu) = -\varepsilon_f\nabla p - F_{pf}^{setII} + \varepsilon_f\nabla.\tau + \rho_f\varepsilon_f g \quad (6)$$

where:

$$F_{pf}^{setII} = \frac{1}{\Delta V}\sum_{i=1}^{n}\left(f_{d,i} + f_i''\right) \quad (7)$$

$$f_{pf,i} = f_{d,i} + f_{\nabla p,i} + f_{\nabla.\tau,i} + f_i''$$

- Set III (Model B):

$$\frac{\partial(\rho_f\varepsilon_f u)}{\partial t} + \nabla.(\rho_f\varepsilon_f uu) = -\nabla p - F_{pf}^{setIII} + \nabla.\tau + \rho_f\varepsilon_f g \quad (8)$$

where:

$$F_{pf}^{setIII} = \frac{1}{\varepsilon_f\Delta V}\sum_{i=1}^{n}\left(f_{d,i} + f_i''\right) - \frac{1}{\Delta V}\sum_{i=1}^{n}\left(\rho_f V_{p,i} g\right) \quad (9)$$

$$f_{pf,i} = \frac{1}{\varepsilon_f}(f_{d,i} + f_i'') - \rho_f V_{p,i} g$$